\documentclass[prd,aps,a4paper,floatfix,amsmath,amssymb,twocolumn]{revtex4-2}
\usepackage[margin=3cm]{geometry}
\usepackage{mathtools}
\usepackage{graphicx}
\usepackage[dvipsnames]{xcolor}
\usepackage[toc,page]{appendix}
\usepackage{array}
\usepackage[pdftex,breaklinks,colorlinks,
linkcolor=Blue,
citecolor=teal,
anchorcolor=red,
urlcolor=cyan]{hyperref}

\newcommand{\bracket}[1]{\left( #1 \right)}

\begin{document}

\title{Critical collapse of a spherically symmetric ultrarelativistic 
	fluid in $2+1$ dimensions}

\author{Patrick Bourg, Carsten Gundlach}
\affiliation{Mathematical Sciences, University of Southampton,
	Southampton SO17 1BJ, United Kingdom}
\date{23 April, revised 17 July, 2021}

\begin{abstract}
	We carry out numerical simulations of the gravitational
	collapse of a perfect fluid with the ultrarelativistic
	equation of state $P=\kappa\rho$, in spherical symmetry in
	$2+1$ spacetime dimensions with $\Lambda<0$. At the threshold
	of prompt collapse, we find type~II critical phenomena
	(apparent horizon mass and maximum curvature scale as powers
	of distance from the threshold) for $\kappa\gtrsim 0.43$, and
	type~I critical phenomena (lifetime scales as logarithm of
	distance from the threshold) for $\kappa\lesssim 0.42$. The
	type~I critical solution is static, while the type~II critical
	solution is not self-similar (as in higher dimensions), but
	contracting quasistatically.
\end{abstract}

\maketitle

\tableofcontents

\section{Introduction}

Since the seminal paper of Choptuik \cite{Choptuik93}, it has become clear
that for many simple, typically spherically symmetric, self-gravitating
systems, such as a scalar field or perfect fluid, the evolutions of
generic initial data close to the threshold of black hole formation
exhibit several universal properties, which are now collectively
called type~II critical phenomena at the threshold of gravitational
collapse. These are interesting in particular as a route to the
formation of naked singularities from regular initial data. (See
Ref.~\cite{GundlachLRR07} for a review).

Consider a one-parameter family of initial data with parameter
$p$. Suppose that there exists a threshold value $p=p_\star$
so that supercritical initial data, $p > p_\star$, eventually
collapses into a black hole, while subcritical initial data, $p <
p_\star$, instead disperses.

In type~II critical phenomena, one observes in the case of
supercritical data that the black-hole mass obeys a power law $M
\propto (p-p_\star)^\delta$, where the exponent $\delta>0$ does not
depend on the initial data. (It does depend on the type of matter,
within certain universality classes). On the other hand, for
subcritical data, it is the maximum curvature that scales,
$\text{Ric}_\text{max} \propto (p_\star-p)^{-2\gamma}$, where $\gamma>0$
is also independent of the initial data.

In $d+1$ dimensions with $d > 2$, owing to the fact that mass has
dimension length$^{d-2}$, the exponents $\delta$ and $\gamma$ are
related to each other via $\delta = (d-2)\gamma$, as can be shown from
dimensional analysis. These properties near the black-hole threshold
are explained through the existence of a \textit{critical solution},
which has the key properties of being regular, self-similar and having
precisely one growing mode. This critical solution appears as an
intermediate attractor in the time evolution of any
near-critical initial data. As we fine-tune to the black-hole
threshold, $p \to p_\star$, the (unique) growing mode is increasingly
suppressed, so that the critical solution persists on arbitrarily small
scales and correspondingly large curvature without collapsing and thus
without an event horizon forming. A naked singularity then forms at
some finite central proper time for exactly critical data $p = p_\star$.

In type~I critical phenomena (by contrast to type~II), the
critical solution is stationary, and instead of mass and curvature
scaling, one observes scaling of the lifetime of its appearance as an
intermediate attractor, $t_p\propto\ln|p-p_\star|$.

Critical phenomena in spherical symmetry have been observed in
numerous matter models. Since black holes are in general characterized
by their mass, charge, and angular momentum, the complete picture of
critical phenomena necessarily requires investigation beyond
spherical symmetry. However, even generalizing spherically symmetric
initial data to axisymmetric ones brings substantial numerical and
analytical complications in $3+1$ and higher dimensions. As a result,
there have been far fewer studies devoted to studying critical
phenomena beyond spherical symmetry. 

For this reason, it may be helpful to investigate critical collapse in
$2+1$ dimensional spacetime as a toy model. In $2+1$ dimensions, all
variables are only functions of time and radius for both spherical symmetry
and axisymmetry. This avoids much of the additional technical
complication of axisymmetry. As in Ref.~\cite{Jalmuzna17}, we call a
solution circularly symmetric if it admits a spacelike Killing vector
$\partial_\theta$ with closed orbits. More specifically, we call it
spherically symmetric if there is no rotation, and we call it
axisymmetric with rotation.

An interesting peculiarity of $2+1$ is that black holes cannot
form without the presence of a negative cosmological constant. This
fact seemingly causes a paradox since, on the one hand, a cosmological
constant is required for black holes to form, and thus for the
possibility of critical phenomena to occur. On the other hand, one
expects any type~II critical solution to not depend on the
cosmological constant, due to the fact that as the critical solution
persists on arbitrarily small length scales, the cosmological constant
is expected to become dynamically irrelevant, and so the underlying
Einstein and fluid equations become approximately scale invariant.
That is probably related at a deep level to the fact that in $2+1$
dimensions the mass is dimensionless and it follows that the usual
argument to relate the two exponents $\delta$ and $\gamma$
fails.

Aside from the present paper, the only studies that have investigated
critical phenomena in $2+1$ dimensions were restricted to the massless
nonrotating \cite{Pretorius00,Jalmuzna15} and rotating
\cite{Jalmuzna17} scalar fields. An interesting fact that emerged from
those studies is that in $2+1$ dimensions, the nonrotating critical
solution for the massless scalar field is \textit{continuously}
self-similar, as opposed to its $3+1$ version, where it is discretely
self-similar. Furthermore, the critical solution is well approximated
inside the past light cone of its singularity by exactly
self-similar solutions to the $\Lambda=0$ Einstein equations. Outside
the light cone it can be approximated by a different $\Lambda=0$
exact solution. This patchwork critical solution has three growing modes,
but it is conjectured that when $\Lambda$ is taken into account
nonperturbatively, the true critical solution is analytic and retains
only the top growing mode. This conjecture is in part supported by the
fact that under this assumption, one can find a scaling law for the
black-hole mass such that $\delta = 2 \gamma / (2 \gamma+1)$, consistent
with the numerical results.

In this paper, we study the spherically symmetric collapse of a
perfect fluid in $2+1$ in anti-de~Sitter (from now, AdS) space with
the linear (ultrarelativistic) equation of state $P = \kappa
\rho$. Although an important motivation for looking at collapse in
$2+1$ dimensions is that axisymmetry with rotation is as simple as
spherical symmetry, we begin in this paper with a study of
spherically symmetric, nonrotating, collapse.

The structure of the paper is as follows. In
Section~\ref{section:section2}, we give a brief description of the
equations we solve and their numerical implementation. We refer the
reader to Ref.~\cite{Carsten21} for a complete discussion and details of
our numerical implementation. In Sec.~\ref{section:section3}, we
present the results of our numerical investigation of the threshold of
prompt collapse for a spherically symmetric perfect fluid in $2+1$
dimensions. We show evidence of both type~I and type~II behavior
depending on the value of $\kappa$. The type~I critical
solution is static, describing a metastable star. The type~II
critical solution shrinks quasistatically, moving adiabatically
through the family of static stars. (A slightly different
approximation is needed in the thin atmosphere of the star, where
the outflow speed is relativistic.) Sec.~\ref{section:conclusions}
contains our conclusions. In appendixes, we show that no regular
continuously self-similar solution exists, review the static
solutions, and show how they relate to the quasistatic solution.

\section{Einstein and Fluid equations in polar-radial coordinates}
\label{section:section2}

We refer the reader to the companion paper \cite{Carsten21} for a complete
discussion. We use units where $c=G=1$.

In spherical symmetry in $2+1$ dimensions, we introduce generalised
polar-radial coordinates as
\begin{eqnarray}
\label{trmetric}
ds^2&=&-\alpha^2(t,r)\,dt^2+a^2(t,r)R'^2(r)\,dr^2 \nonumber \\ &&
+ R^2(r)d\theta^2. 
\end{eqnarray}
Note that our choice $g_{rr}=a^2R'^2$ makes $a$ invariant under a
redefinition of the radial coordinate, $r\to \tilde r(r)$. 

We impose the gauge condition $\alpha(t,0)=1$ ($t$ is proper
time at the center), and the regularity condition $a(t,0)=1$
(no conical singularity at the center). The gauge is fully
specified only after specifying the function $R(r)$. In our numerical
simulations we use the compactified coordinate
\begin{equation}
\label{compactifiedR}
R(r)=\ell\tan(r/\ell),
\end{equation}
with different values of the cosmological scale $\ell$ defined by
\begin{equation}
\ell:={1 \over \sqrt{-\Lambda}},
\end{equation}
but for clarity we write $R$ and $R'$ rather than the explicit expressions.

In our coordinates, the Misner-Sharp mass $M$ is given by
\begin{eqnarray}
\label{Mdefcoords}
M(t,r)&:=&{R^2\over\ell^2}-{1\over a^2}.
\end{eqnarray}

The stress-energy tensor for a perfect fluid is 
\begin{equation}
T_{ab}=(\rho+P)u_{a}u_{b}+Pg_{ab},
\end{equation}
where $u^a$ is tangential to the fluid worldlines, with $u^a u_a=-1$, and $P$
and $\rho$ are the pressure and total energy density measured in the fluid frame.
In the following, we assume the one-parameter family of ultrarelativistic fluid
equations of state $P=\kappa\rho$, where $0<\kappa<1$. 

The 3-velocity is decomposed as
\begin{equation}
u^\mu=\{u^t,u^r,u^\theta\}=\Gamma\left\{\frac{1}{\alpha},\frac{v}{aR'},
0 \right\},
\end{equation}
where $v$ is the physical velocity of the fluid relative to observers at
constant $R$, with $-1<v<1$, and 
\begin{equation}
\Gamma:=\left(1-v^2\right)^{-1/2}
\end{equation}
is the corresponding Lorentz factor.

The stress-energy conservation law $\nabla_a T^{ab}=0$, which
together with the equation of state governs the fluid evolution,
can be written in balance law form
\begin{equation}
{\bf q}_{,t}+{\bf f}_{,r}={\bf S}, \label{2p1:XY_cons_law}
\end{equation}
where we have defined the conserved quantities
\begin{equation}
{\bf q}:=\{\Omega,Y\}
\end{equation}
given by
\begin{eqnarray}
\Omega&:=& R'R \tau,\\
Y&:=& R'v \sigma,
\end{eqnarray}
the corresponding fluxes ${\bf f}$ given by
\begin{eqnarray}
f_{(\Omega)}&:=& {\alpha\over a} R v \sigma,\\
f_{(Y)}&:=& {\alpha\over a}(P+v^2\sigma),
\end{eqnarray}
the corresponding sources ${\bf S}$ given by 
\begin{eqnarray}
S_{(\Omega)}&:=& 0, \\
S_{(Y)}&:=& a \alpha R R' \Bigl[-\frac{v^2 \sigma}{a^2 R^2} +
2 P (8 \pi P-\Lambda ) \nonumber \\ && 
-\sigma\left(1-v^2\right) (16 \pi P-\Lambda) \Bigr],
\label{SOmega}
\end{eqnarray}
and the shorthands
\begin{eqnarray}
\label{sigmadef}
\sigma&:=&\Gamma ^2 (1+\kappa) \rho, \\
P&:=&\kappa\rho, \\
\tau&:=&\sigma-P.
\label{taudef}
\end{eqnarray}
In Eq.~\eqref{SOmega}, we have already used some of the Einstein equations
to express metric derivatives in terms of stress-energy terms.

At any given time, the balance laws [Eq.~\eqref{2p1:XY_cons_law}] are used
to compute time derivatives of the conserved quantities $\bf
q$, using standard high-resolution shock-capturing methods. The
${\bf q}$'s are evolved to the next time step via a second-order
Runge-Kutta step. At each (sub-)time step, the metric variables
are then updated through the Einstein equations
\begin{eqnarray}
\label{dlnalphaaoRplphadr}
(\ln \alpha a)_{,r}&=&8\pi a^2RR'(1+v^2)\sigma, \\
M_{,r} &=& 16 \pi \Omega.
\label{dMdr}
\end{eqnarray}

Our numerical scheme is totally constrained, in the sense
that only the matter is updated through evolution equations.
Our numerical scheme exploits this to make
$\Omega$ and $M$ exactly conserved in the discretized equations.

Another useful Einstein equation, compatible with the above
ones via stress-energy conservation, is
\begin{equation}
\label{Mteqn}
M_{,t}=-16\pi f_{(\Omega)}.
\end{equation}

\section{Numerical results}
\label{section:section3}

\subsection{Initial data}

The numerical grid is equally spaced in the compactified coordinate $r$, as
defined in Eq.~\eqref{compactifiedR}, with $800$ grid points, and for all
values of $\Lambda$ its outer boundary is fixed at the same area radius $R$.
For reasons that will be made clear, we fix, unless
otherwise stated, $R_\text{max} \simeq 1.25$ for $\kappa \geq 0.43$ and
$R_\text{max} \simeq 10$ for $\kappa \leq 0.42$.

We choose to initialize the intermediate fluid variables 
\begin{equation}
\omega:={\Omega\over R'R}, \qquad \eta:={Y\over R'R}
\end{equation}
as double Gaussians in $R$,
\begin{align}
\omega(0,R) &= \frac{p_\omega}{2} \bracket{e^{-\bracket{
			\frac{R-R_\omega}{\sigma_\omega}}^2} +
	e^{-\bracket{\frac{R+R_\omega}{\sigma_\omega}}^2}}, 
\label{2p1:initialdata_gaussian} \\
\eta(0,R) &= \frac{p_\eta}{2} \bracket{e^{-\bracket{
			\frac{R-R_\eta}{\sigma_\eta}}^2} + e^{-\bracket{
			\frac{R+R_\eta}{\sigma_\eta}}^2}}.
\label{2p1:gaussian_initial_data}
\end{align}
where $p_\omega$ and $p_\eta$ are the magnitudes, $R_\omega$ and
$R_\eta$ are the displacements from the center, and $\sigma_\omega$ and
$\sigma_\eta$ are the widths of the Gaussians. Note that $p_\omega$ has
dimension length$^{-2}$, while $p_\eta$ is dimensionless.

For $\kappa \geq 0.43$, we fix $\sigma_\omega = 0.2$ and
$\sigma_\eta = 0.15$ and consider three types of initial data:

1) Time-symmetric off-centered: $p_\eta = 0$, $R_\omega = 0.4$, 

2) Time-symmetric centered: $p_\eta = 0$, $R_\omega = 0$, and 

3) Initially ingoing off-centered: $p_\eta = -0.2$, $R_\omega = R_\eta = 0.4$.

For $\kappa \leq 0.42$, we consider time-symmetric off-centered and ingoing
initial data as given above, and time-symmetric centered $R_\omega = 0$,
$\sigma_\omega = 0.05$, $p_\eta = 0$, and $\sigma_\eta = 0.15$.

In all cases, the remaining parameter $p_\omega = : p$ is fine-tuned
to the black-hole threshold.

The space of initial data parametrized by $p$ can be subdivided into
four regions with boundaries $p_- < p_\star< p_+$ as follows: At
$p=p_-$, the total mass is zero, $M_\infty=0$, while $p=p_\star$
corresponds to the critical value separating subcritical (initially
dispersing) from supercritical (promptly collapsing) initial
data. Finally, $p_+$ is defined so that a trapped surface,
characterized by $(\nabla R)^2 = 0$, is already present for initial
data with $p > p_+$.

\begin{table} \centering
	\begin{tabular}{{p{0.5\columnwidth}>{\centering}
		p{0.15\columnwidth}>{\centering}p{0.15\columnwidth}
		>{\centering\arraybackslash}p{0.1\columnwidth}}}
		\hline \hline
		Initial data ($\kappa=0.5$) & $p_-$ & $p_\star$ & $p_+$ \\
		\hline
		Off-centered, $\tilde{\mu}=0.01$, & 0.280 & 0.309 & 0.324 \\
		Off-centered, $\tilde{\mu}=0.1$,  & 0.280 & 0.402 & 0.612 \\
		Centered,     $\tilde{\mu}=0.1$,  & 0.995 & 1.174 & 1.354 \\
		Ingoing,      $\tilde{\mu}=0.1$,  & 0.280 & 0.377 & 0.612 \\
		Off-centered, $\tilde{\mu}=1$,    & 0.280 & 0.531 & 3.087 \\
		Off-centered, $\tilde{\mu}=10$,   & 0.280 & 0.572 & 27.33 \\
		\hline \hline
	\end{tabular}
	\caption{The relation between $p_-$, $p_\star$, $p_+$, and the
		initial data that we are considering in this paper for
		$\kappa = 0.5$.}
	\label{table:pminus_star_plus}
\end{table}

In $2+1$, black-hole solutions with $M_\infty > 0$ are separated from
the vacuum AdS solution $M_\infty=-1$ by a mass gap \cite{Banados92},
so that no initial data with $p < p_-$ can collapse into a black hole.

As already stated, in $2+1$ dimensions, a negative cosmological constant is
necessary for the formation of black holes, and thus for critical phenomena
to occur. Since the cosmological constant introduces a length scale $\ell$
into the system, we need to consider different sizes of the initial
data with respect to $\ell$, which can be quantified by considering the
dimensionless quantity
\begin{equation}
\tilde{\mu} := -\Lambda \sigma_\omega^2 = 
\bracket{\frac{\sigma_\omega}{\ell}}^2.
\label{2p1_crit:tildemu}
\end{equation}
$\sigma_\omega$ is kept fixed as given before, while we vary
$\tilde{\mu}$ and thus $\ell$. For $\kappa \geq 0.43$, we set
$\tilde{\mu}=0.1$. For $\kappa=0.5$, we in addition study the cases
$\tilde{\mu} = 0.01$, $1$ and $10$, corresponding to a range of values
of the cosmological length scale that are ``small'' to ``large''
compared to the initial data. For $\kappa \leq 0.42$, we set
$\Lambda=-\pi^2/4$ ($\tilde{\mu} \simeq 0.006$).

In Table~\ref{table:pminus_star_plus}, we record the values of $p_-$,
$p_\star$ and $p_+$ for different families of initial data with
$\kappa=0.5$. Note that, unlike in higher dimensions, $p_- \to p_+$
as $\Lambda \to 0$.

Regularity at the timelike outer boundary of spacetimes with
$\Lambda<0$ does not allow a stress-energy flux through it. For a
scalar field, this enforces homogeneous Dirichlet (reflecting)
boundary conditions, whereas for perfect fluid matter its energy
density needs to vanish at the boundary. Dynamically, this is
enforced by an inward Hubble acceleration of the matter, so that any
test particles on timelike geodesics, at least, must turn around
inwards. Hence, it is \textit{a priori} possible for data to collapse only
after being reflected, possibly several times, from the boundary, as
was observed for the massless scalar field in Ref.~\cite{Bizon11}. 

As we impose (unphysical) outer boundary conditions at a finite
radius, most of the energy that is outgoing in fact leaves the
numerical domain. Thus, we cannot directly investigate the
reflective property of AdS here. In this sense, we fine-tune to the
threshold of prompt collapse. 

Independently, because of our polar time
slices, our code stops when a trapped surface first appears on a time
slice, and so we cannot obtain the final black-hole mass, so in this
sense we measure the mass of the apparent horizon when it first
touches a polar time slice. (However, it is likely that given enough
time, all matter eventually falls into the black hole, so the black-hole
mass becomes equal to the total mass $M_\infty$.)

In the following, we are interested in initial data where $p \simeq p_\star$
and we refer to ``sub$n$'' data as subcritical data for which $\log_{10}
(p_\star-p) \simeq -n$, and to ``super$n$'' as supercritical data
with $\log_{10} (p-p_\star) \simeq -n$.

\subsection{Overview of results}

For the equation of state $P=\kappa\rho$ with $\kappa\lesssim 0.42$,
we find type~I critical phenomena: time evolutions of initial data
near the black-hole threshold approach a static solution before either
collapsing to a black hole or dispersing, and the time this
intermediate attractor is seen for scales as $t_p\sim -\ln|p-p_\star|$.
The static type~I critical solution is not universal.

For $\kappa\gtrsim 0.43$ we observe scaling of the apparent horizon
mass and maximum curvature as powers of distance to the threshold of
(prompt) collapse, characteristic of type~II critical
phenomena. The type~II critical solution is universal but not self-similar.
It is instead quasistatic, running adiabatically through the one-parameter
family of regular static solutions.

\subsection{$\kappa\lesssim 0.42$: Type~I critical collapse}

\subsubsection{Lifetime scaling}

In type~I critical phenomena, the critical solution is stationary or
time-periodic instead of self-similar. Furthermore, there
is a nonvanishing mass gap at the black-hole threshold, and so the critical
solution is usually thought of as a metastable star.

\begin{figure*}
	\includegraphics[width=2.\columnwidth, height=1.3\columnwidth]
	{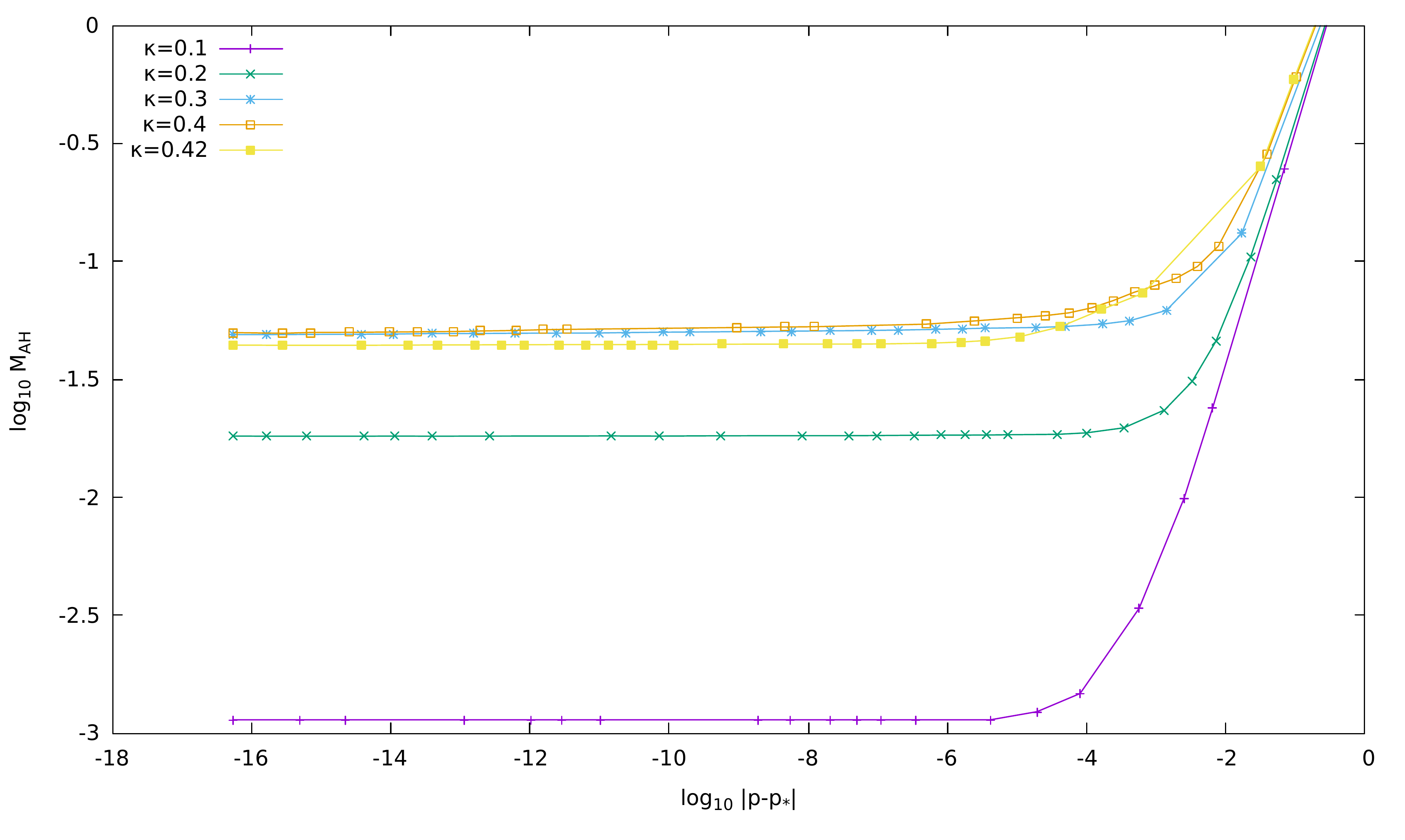}
	\caption{Apparent horizon mass for different values of $\kappa
		\leq 0.42$. The mass does not scale, and there is a mass gap
		at the black-hole threshold instead. We similarly find an
		upper bound for the maximum curvature.}
	\label{fig:Mscale_typeI}
\end{figure*}

In Fig.~\ref{fig:Mscale_typeI}, we plot the apparent horizon mass
$M_\text{AH}$ against $p-p_\star$ for different values of
$\kappa$. For $\kappa \leq 0.42$, we find the existence of a mass gap,
corresponding to type~I critical phenomena. Similarly, the maximum
curvature is bounded above, but we choose not to show it here in order
to avoid cluttering. This plot was obtained using a
second-order limiter. With a first-order (Godunov) flux
limiter, the mass scales, but it does so in a step-size manner.
We believe that this is an artifact of the Godunov limiter.

To further investigate the type~I behavior, we need to define a
measure of the length scale of the solution. We can do this by recording,
for example, the central density and mass at the outer boundary, defined by
\begin{equation}
\rho_0(t):= \rho(t,0), \quad M_\text{OB}(t) := M(t,R_\text{max})
\end{equation}
and the radius where the mass vanishes,
\begin{equation}
M(t,R_M(t)) := 0. \label{rho0_MOB_RM}
\end{equation}

\begin{figure*}
	\includegraphics[width=1.02\columnwidth, height=1.\columnwidth]
	{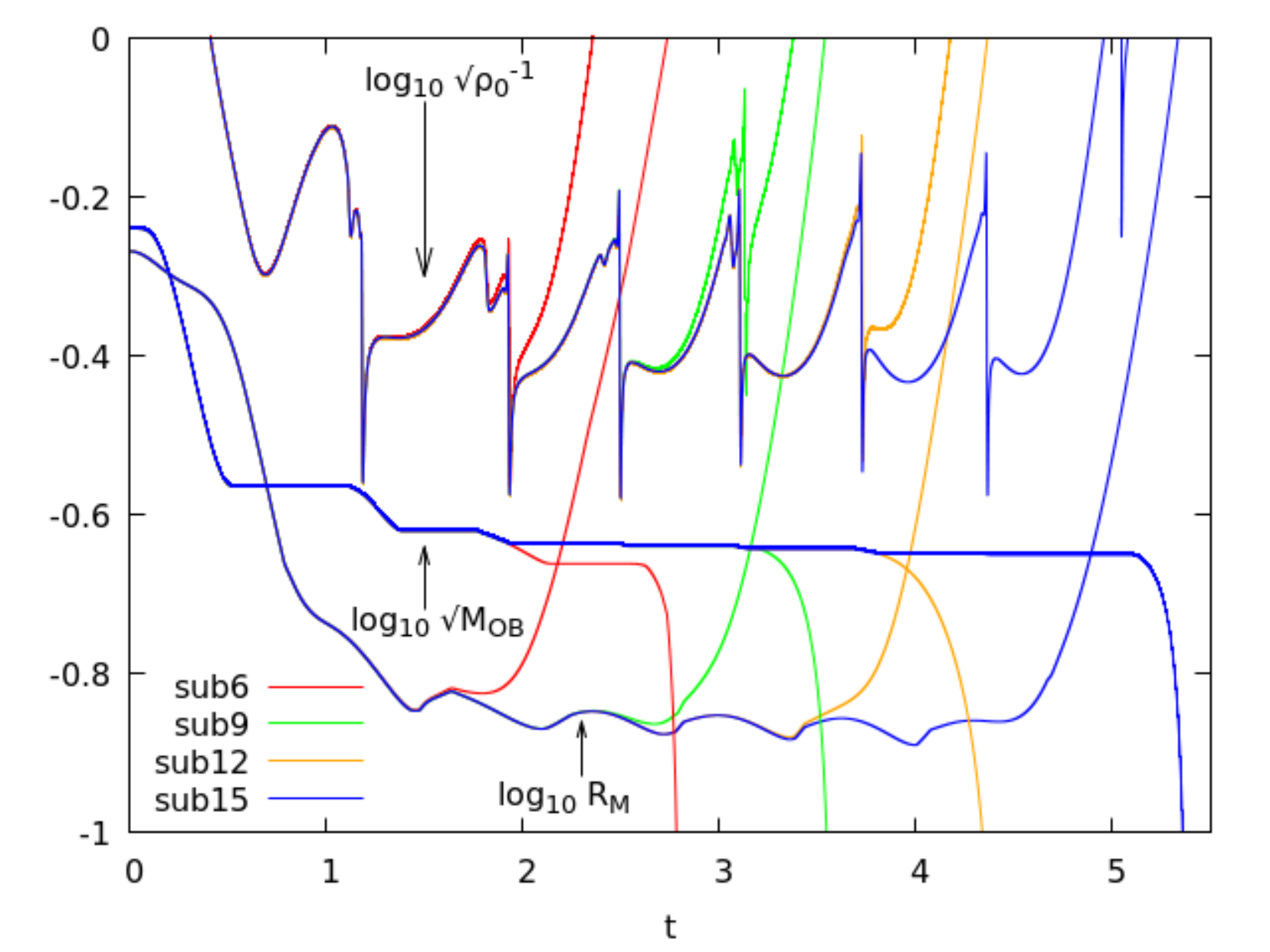}
	\includegraphics[width=1.02\columnwidth, height=1.\columnwidth]
	{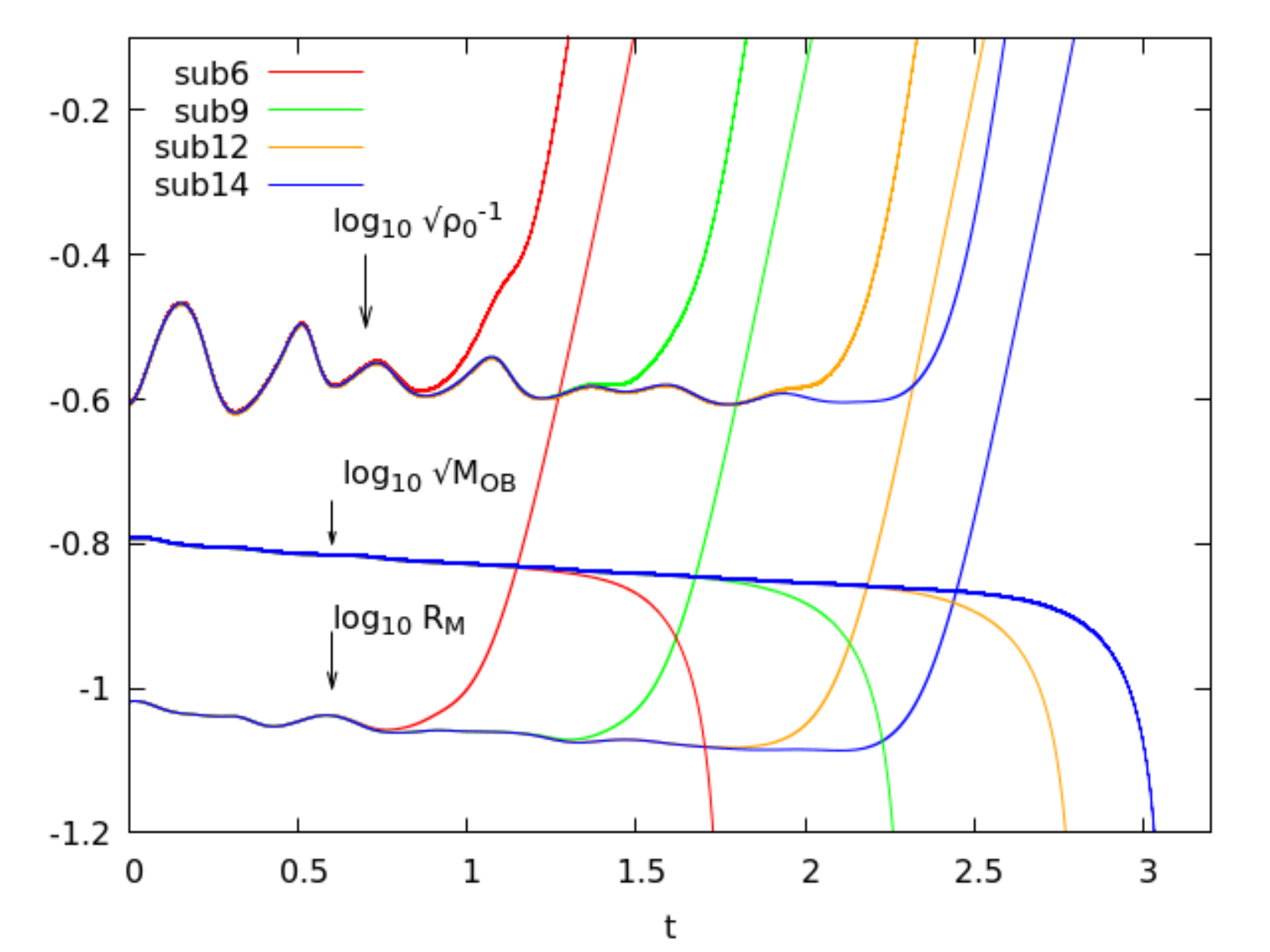}
	\caption{Linear-log-plot of $R_M(t)$, $\sqrt{\rho_0^{-1}(t)}$, and
		$\sqrt{M_\text{OB}(t)}$ for the evolution of sub6-sub15
		initial data with $\kappa=0.4$. The plot on the left shows
		off-centered initial data, evolved with the MC limiter.
		The plot on the right shows more compact and
		centered initial data, evolved with the Godunov	limiter.}
	\label{fig:rho0rmmobsubkappa04}
\end{figure*}

In Fig.~\ref{fig:rho0rmmobsubkappa04}, we plot $\sqrt{\rho_0^{-1}}$,
$\sqrt{M_\text{OB}}$, and $R_M$ for off-centered (left) and
centered (right) initial data with $\kappa=0.4$, evolved with the
monotonized central (MC) and Godunov limiters, respectively. Both sets of
results are shown at different levels of fine-tuning.

In both cases, we see that $R_M$ and $M_\text{OB}$ are approximately
constant during the critical regime. In the former case, the
central density is subject to the apparition of periodic shocks, which
cause the periodic structure for $R_M$ and $\rho_0$. In the latter
case, the periodic shocks are not present and the density converges to
some finite value. We find that in this case $M_\text{OB}$ and $R_M$
evolve slightly. This is due to the fact that the Godunov
limiter is known to introduce a great deal of numerical diffusion
\cite{Leveque02}. A benefit of this diffusion is that the aforementioned
shocks, developing at the outer boundary, are not present, and the density
converges to some finite value. In both cases, however, less
fine-tuned initial data peel off from the critical behavior faster
than more fine-tuned data. This suggests that the critical solution
has a single growing mode which is being progressively suppressed as
we fine-tune to the black-hole threshold.

The shocks mentioned above in the case of the second-order limiter
originate from the unphysical boundary conditions imposed at the
numerical outer boundary. Although one might wonder how these shocks
interact with the behavior of the critical solution, one can still
reasonably believe that the type~I phenomena are not a numerical
artifact. One reason is that we still find type~I phenomena in
the second set of simulations described above, where the
aforementioned instabilities do not occur.

As the critical solution does not depend on $t$, for its linear
perturbations we can make the ansatz
\begin{equation}
\delta Z(t,x) = \sum_{i=0}^{\infty} C_i(p)\, e^{\sigma_i {t \over \ell}}\, Z_i(x),
\label{crit_lin_pert_typeI}
\end{equation}
where $Z$ stands for any dimensionless metric or matter variable.

By definition, the critical solution has a single growing mode,
${\rm Re}\, \sigma_0 > 0$. Since the solution is exactly critical at
$p=p_\star$, this implies that $C_0(p) \sim p-p_\star$.

We define the time $t = t_p$ to be the time where the growing perturbation
becomes nonlinear. We can take this to be
\begin{equation}
(p-p_\star) e^{\sigma_0 {t_p \over \ell}} \simeq 1,
\end{equation}
and so
\begin{equation}
\label{lifetimescaling}
t_p = \frac{\ell}{\sigma_0} \ln|p-p_\star| + \text{constant}.
\end{equation}

The exponent $\sigma_0$ for $\kappa=0.4$, for example, can be read off
from Fig.~\ref{fig:rho0rmmobsubkappa04}. Specifically, we treat the
value of $p$ of our best fine-tuned data as a proxy for $p_\star$. We
then record, as a function of $p-p_\star$ from sub8 to sub15
initial data, the time $t=t_p$ where, say $R_M$, peels off. We
similarly compute $\sigma_0$ for other values of $\kappa$.

In Table~\ref{table:nu_sigma0_typeI}, we show $\sigma_0$ for different
values of $\kappa$ and find that $\sigma_0$ increases approximately linearly
with $\kappa$.

\begin{table} \centering
	\begin{tabular}{{>{\centering}p{0.45\columnwidth}>{\centering\arraybackslash}p{0.45\columnwidth}}}
		\hline \hline
		$\kappa$ & $\sigma_0$ \\
		\hline
		$0.30$ & 4.97 \\
		$0.32$ & 5.54 \\
		$0.34$ & 5.89 \\
		$0.36$ & 6.46 \\
		$0.38$ & 7.19 \\
		$0.40$ & 8.84 \\
		$0.42$ & 9.70 \\
		\hline \hline
	\end{tabular}
	\caption{The value of $\sigma_0$ as a function of $\kappa$. We
		have obtained $\sigma_0$ from the lifetime scaling
		[Eq.~\eqref{lifetimescaling}] of the critical solution.}
	\label{table:nu_sigma0_typeI}
\end{table}

\subsubsection{The critical solution}

All spherically symmetric static solutions (and in fact all
rigidly rotating axisymmetric stationary solutions) with
$\Lambda \leq 0$ were found in Ref.~\cite{Cataldo04}, for arbitrary
fluid equations of state.

In Ref.~\cite{Carsten20}, we highlighted the
existence of a two-parameter family of rigidly rotating static star
solutions, for any causal equation of state, which are
analytic everywhere including at the center, and have finite total
mass $M$ and angular momentum $J$. In particular, there is a
one-parameter family of static solutions with a regular center and
finite total mass, parameterised by an overall length scale $s$, see
Appendix~\ref{appendix:appendixB} for a summary of the notation and results
for the specific equation of state $P=\kappa\rho$.

The equation of state $P=\kappa\rho$ itself is scale invariant,
so in the absence of a cosmological constant, the dimensionless quantities
\begin{equation}
Z:=\{R^2\rho,M,\alpha\}
\end{equation}
characterizing a static solution can then only depend on $R/s$.
However, the cosmological constant breaks scale invariance and so the
family of static solutions instead takes the form
\begin{equation}
\label{Zstatic}
Z = \check Z\left({R \over s}, -\Lambda s^2\right),
\end{equation}
where $\check Z$ is the corresponding exact static solution.

In what follows, all the quantities referring to the static solution
have a check symbol, as in Eq.~\eqref{Zstatic}.

For $\Lambda<0$, these stars have finite total mass (and to be a
critical solution, the total mass needs to be positive), but the
density $\rho$ vanishes only asymptotically at infinity. For small
$\mu:=-\Lambda s^2$, the star has an approximate surface [Eq.~\eqref{yc}]
at $R\simeq s \check x_c$, separating the star
proper from a thin atmosphere with negligible self-gravity. (We note
that in the singular limit $\Lambda=0$, the atmosphere disappears
completely. The exterior solution is now vacuum with $M=0$, and in
particular the spatial geometry is a cylinder of constant radius.)

\begin{figure}
	\includegraphics[width=1.\linewidth]{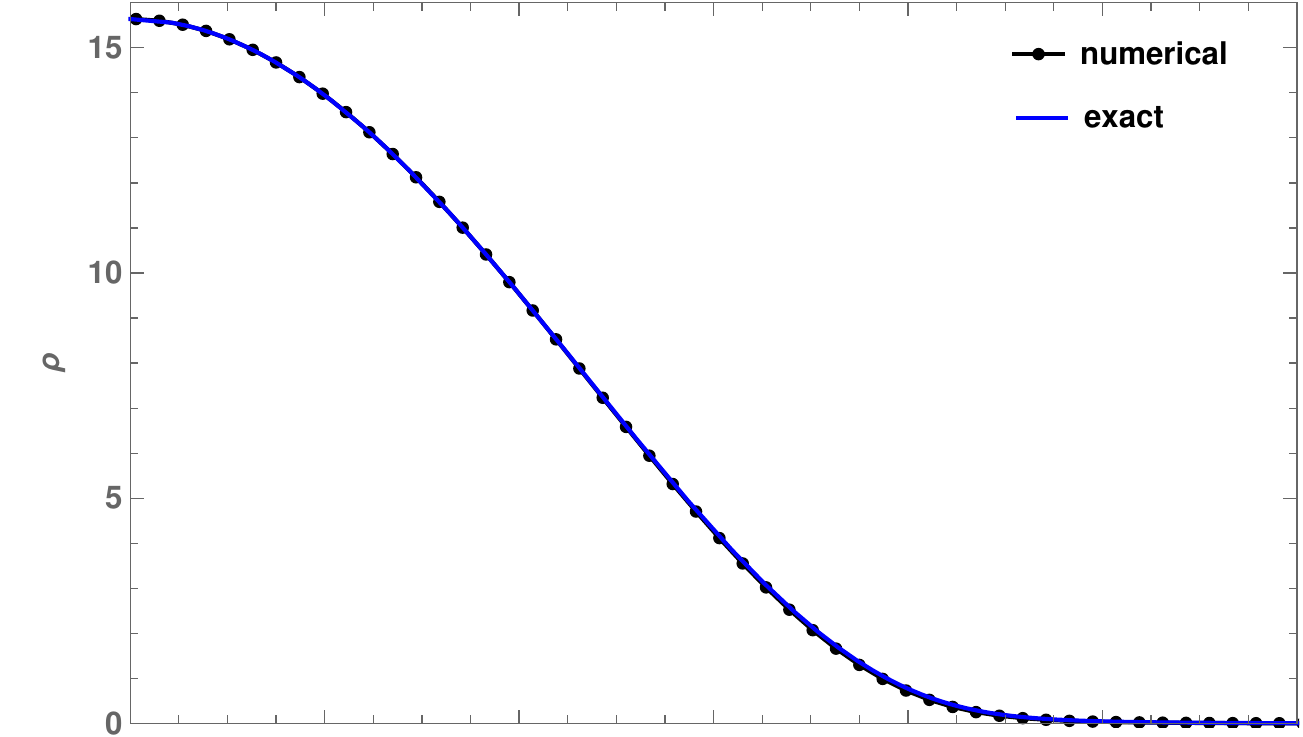}
	\includegraphics[width=1.\linewidth]{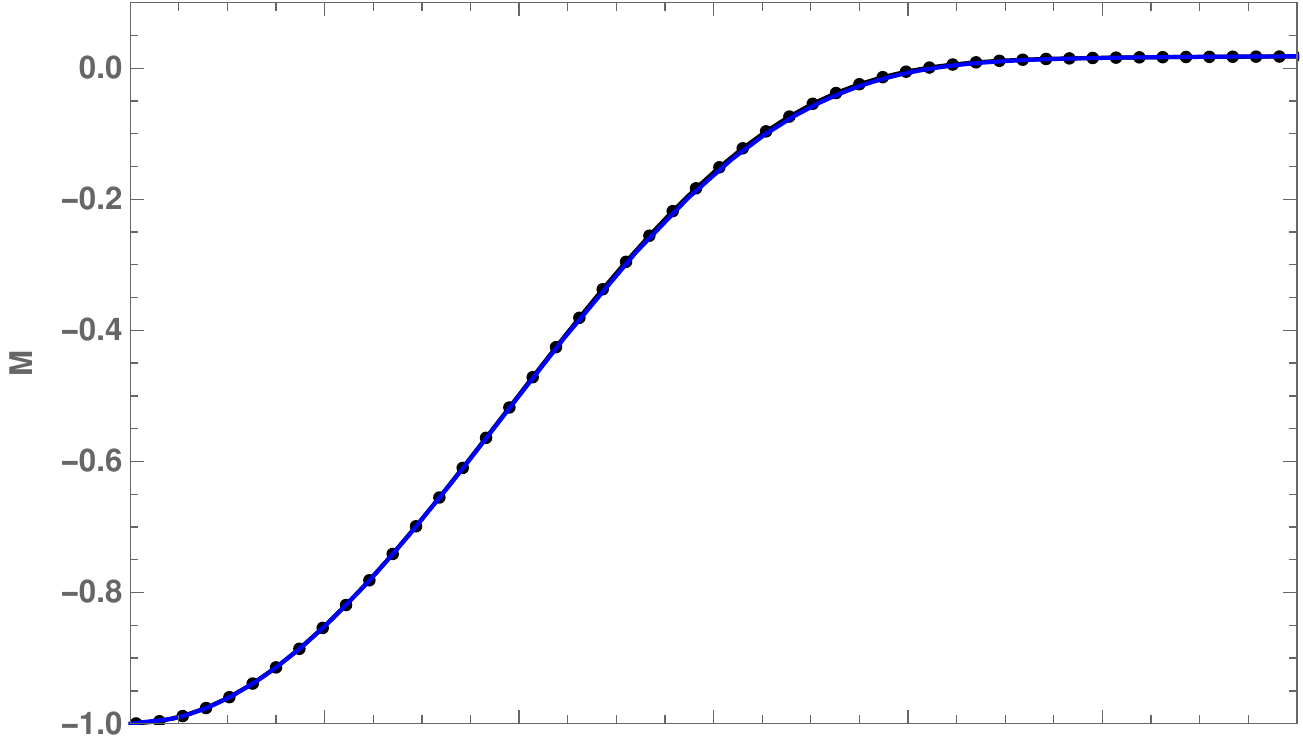}
	\includegraphics[width=1.\linewidth]{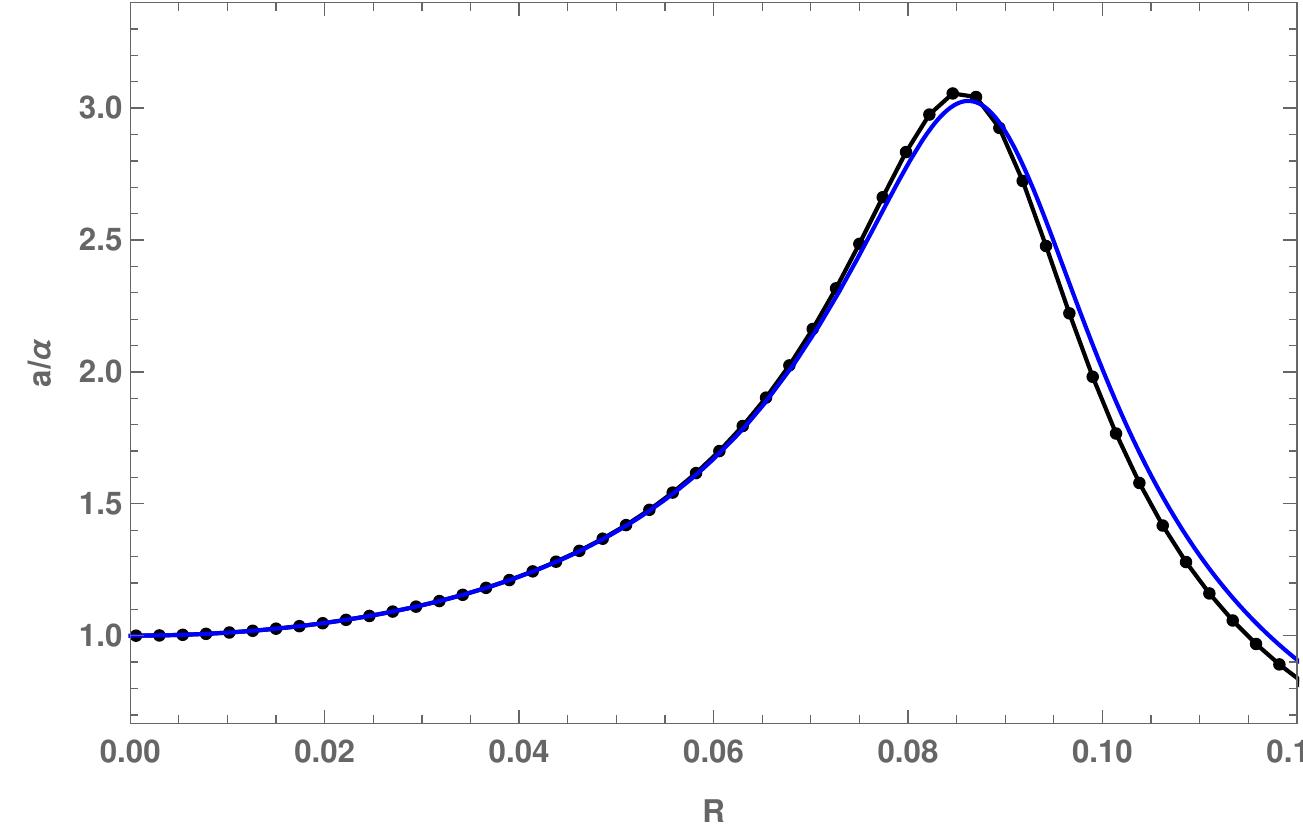}
	\caption{Plots of $M$, $\rho$, and $a/\alpha$ for sub14 centered
		initial data evolved with the Godunov limiter (black) with
		$\kappa=0.4$ at $t \simeq 2$. This is compared with the
		exact static solution (blue), matched using
		Eq.~\eqref{2p1:num_exact_fit_criteria}.}
	\label{fig:kappa04_crit_sol}
\end{figure}

In Fig.~\ref{fig:kappa04_crit_sol}, we provide some evidence that the
critical solution is related to this family of static solution by
plotting our best subcritical solution (with the Godunov limiter) at
time $t \simeq 2$. This is compared to a member of the static
solution, selected to satisfy the condition
\begin{equation}
\Lambda = 8 \pi \kappa \rho_0(t) - \frac{1}{s^2(t)}.
\label{2p1:num_exact_fit_criteria}
\end{equation}
at that time. This ensures that the central density is the same
in the numerical and exact static solutions. We find reasonable
agreement between the numerical and exact static solutions out
to the surface and slightly beyond.

This can also be seen as a further consistency check that the
type~I phenomena are not a numerical artifact as it is otherwise
unlikely that the numerical solution approaches an exact solution to
the Einstein equations.

We find that the critical solution has different masses for the
off-centered, centered and ingoing families of initial data, so
clearly the type~I critical solution is not universal.

\subsection{$\kappa\gtrsim 0.43$: Type~II critical collapse}

\subsubsection{Curvature and mass scaling}

In type~II critical collapse, in a region near the center,
curvature becomes arbitrarily large as the solution approaches a
critical solution with the following defining properties: it is
regular, universal with respect to the initial data, shrinking, and it
has precisely one unstable mode. 

In spherical symmetry, and assuming a continuous, rather than
discrete, scaling symmetry, there exists some adapted coordinate $x
= R/s(t)$, for some function $s(t)$, where $t$ is central proper time
such that a vector $Z(t,x)$ of suitably scaled variables that
characterizes a circularly symmetric solution of the Einstein and
matter equations, is only a function of $x$, $Z(t,x) = Z_\star(x)$.

Since the existence of such a solution is a consequence of the
(approximate) scale invariance of the underlying Einstein and matter
equations, one would expect any quantity of dimension length$^n$ to
scale as $s(t)^n$. In particular, in $d+1$ spacetime dimensions, one
would expect the maximum curvature (and apparent horizon mass) to scale
as 
\begin{equation}
\label{RicMAH_scale}
\text{Ric}_\text{max} \sim s(t_\#)^{-2}, \quad M_\text{AH} \sim s(t_\#)^{d-2}
\end{equation}
where $s(t_\#)$ is the smallest scale the solution reaches before
either dispersing or forming an apparent horizon.

As the critical solution is independent of $t$, for its linear
perturbations we can make the ansatz
\begin{equation}
\delta Z(t,x) = \sum_{i=0}^{\infty} C_i(p)\, s(t)^{-\lambda_i}\, Z_i(x).
\label{crit_lin_pert} 
\end{equation}

We define $t=t_\star$ to be the time where
$s(t_\star) = 0$. Since by definition the critical solution has a single
growing mode, ${\rm Re}\, \lambda_0 > 0$, near the critical time
$t \lesssim t_\star$, all other (decaying, ${\rm Re}\, \lambda_i<0$) modes
are negligible, and we therefore need the fact that $C_0(p) \sim p-p_\star$.

The time $t = t_\#$ when the growing perturbation becomes nonlinear
occurs when $(p-p_\star) s(t_\#)^{-\lambda_0} \sim \mathcal{O}(1)$.
Together with Eq.~\eqref{RicMAH_scale}, one then deduces $\rho_\text{max}$
and $M_\text{AH}$ scale according to the power laws
\begin{align}
M_\text{AH} &= c_M |p-p_\star|^\delta, \label{2p1_crit:mass_scale}\\
\text{Ric}_\text{max} \sim \rho_\text{max} &= 
-\Lambda c_\rho |p_\star -p|^{-2\gamma},
\label{2p1_crit:curvature_scale}
\end{align}
where $c_M$ and $c_\rho$ are dimensionless constants, $\gamma =
1/\lambda_0$, and in space dimension $d \geq 3$, $\delta = (d-2) \gamma$.

We have here slightly generalized the discussion in
Ref.~\cite{GundlachLRR07}, where $s(t)\propto t_\star-t$, with $t$ as the
proper time at the origin, because the critical solution is continuously
self-similar (homothetic). We will see that the critical solution for type II
critical fluid collapse in 2+1 dimensions is not self-similar, but the
generalized discussion still applies.

In $d = 2$, the mass scaling and the value of $\delta$ cannot
be derived by the simple dimensional analysis outlined above. In
Ref.~\cite{Pretorius00}, Pretorius and Choptuik proposed that from the
expression \eqref{Mdefcoords} for the Misner-Sharp mass, one has
$M_\text{AH} = -\Lambda R_\text{AH}^2$, and so one would expect
$M_\text{AH} \sim s(t_\#)^2$. Furthermore, from the dimension of the
curvature, one also expects $\rho_\text{max}^{-1} \sim s(t_\#)^2$,
which, combined with the previous expression, implies
\begin{equation}
M_\text{AH} \sim -\Lambda\rho_\text{max}^{-1},
\label{2p1:MAH_rhomax}
\end{equation}
or $\delta=2\gamma$.
Although we will see that such a relation holds in the present case,
the explanation should explicitly depend on the matter field under
consideration, since a different relation has been shown to hold for
the massless scalar field \cite{Jalmuzna15}, where it was shown that
$R_\text{AH}$ does not scale as suggested by its dimension.

In our numerical investigation of type II critical collapse,
we focus on the equation of state with $\kappa=0.5$, where we
have investigated possible critical behavior by bisecting between
subcritical and supercritical data. Due to the small values of
$\delta$ and $\gamma$ (as compared to critical fluid collapse in 3+1
dimensions \cite{Evans94}), we observe scaling all the way down to
$\log_{10} |p-p_\star| \simeq -15$ in double precision, even at fairly
low numerical resolution and without mesh refinement, as the range of
length scales is not large. We make use of this, or compensate for it,
by working in quadruple precision, even at fairly low grid
resolution. We can then fine-tune to about $\log_{10} |p-p_\star|
\simeq -25$ before we lose resolution (without mesh refinement).

We observe that for $\tilde{\mu} = 0.01$, $0.1$, $1$, and $10$, as we
fine-tune to the black-hole threshold, $M_\text{AH}$ becomes
arbitrarily small while $\rho_\text{max}$ becomes arbitrarily
large. Moreover, both quantities scale so that their product is
constant,
\begin{equation}
\frac{\rho_\text{max} M_\text{AH}}{-\Lambda} \simeq \mathcal{C},
\label{2p1:MAHrhomax_relation}
\end{equation}
where $\mathcal{C}$ is a dimensionless constant independent of $\Lambda$.

\begin{figure*}
	\includegraphics[width=2.\columnwidth, height=1.3\columnwidth]
	{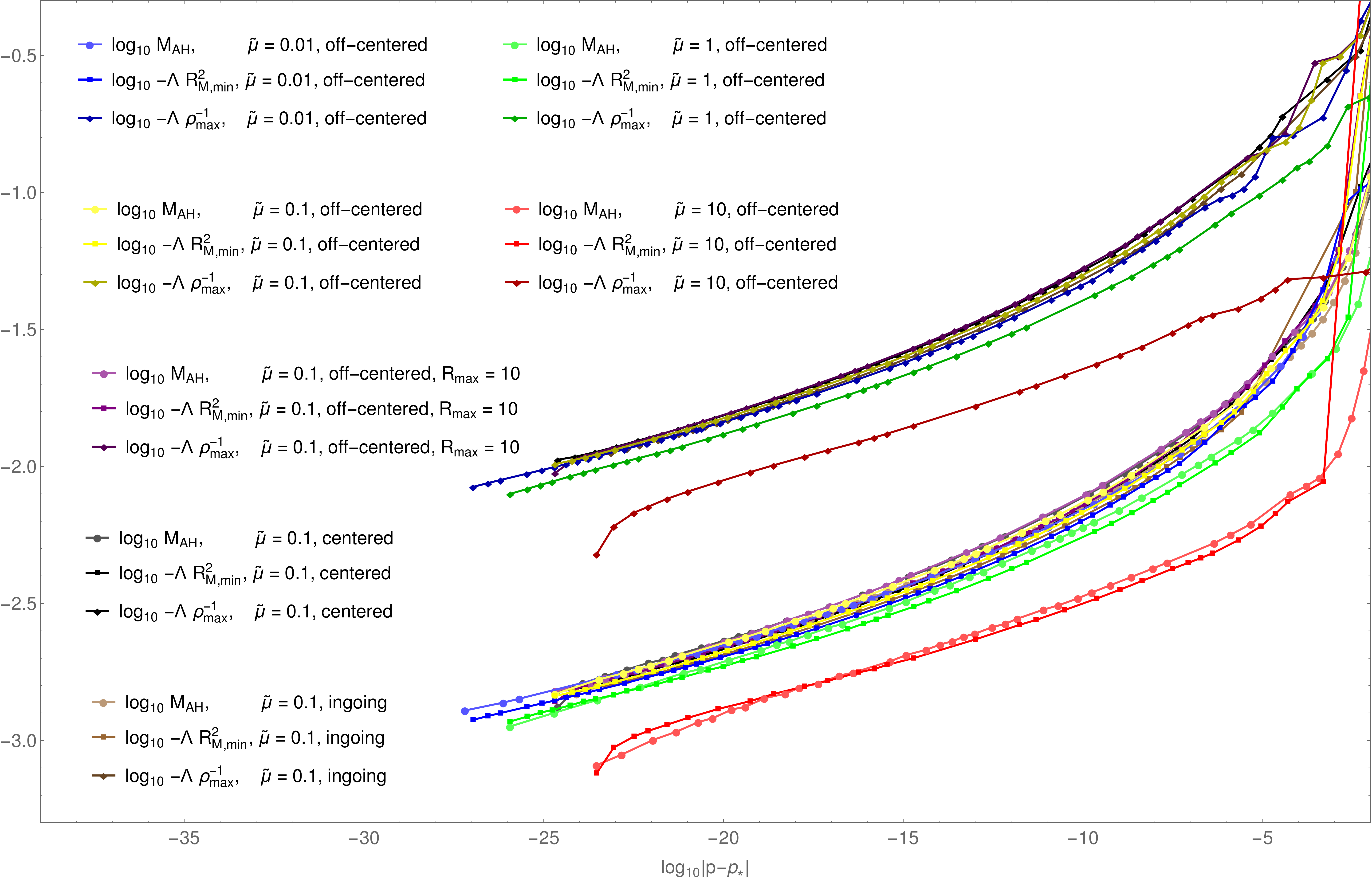}
	\caption{Log-log plot demonstrating the power-law scaling of
		$\rho_\text{max}$ (upper group of curves), $M_\text{AH}$ and
		$R_{M\text{,min}}$ (lower group of curves) for different values of the cosmological constant, different outer boundary locations, and different initial data. The plot gives evidence that the relations
		$-\Lambda \rho_\text{max}^{-1} \sim M_\text{AH} \simeq -\Lambda
		R^2_{M\text{,min}}$ hold, are universal, and are independent of the
		cosmological constant. The constants $c_M$, $c_\rho$, and $c_R$ in general
		depend on the family of initial data and $\tilde\mu$, but here are almost
		universal except for $\tilde\mu=10$ (curves somewhat below each group).}
	\label{fig:2d:2p1:mass_ricci_scaling}
\end{figure*}

Furthermore, we empirically observe that for sub15 data
onwards, $M_\text{AH}$ and $\rho_\text{max}$ are well fitted by
power laws \eqref{2p1_crit:mass_scale} and
\eqref{2p1_crit:curvature_scale}, respectively, where the exponents are
related by
\begin{equation}
\delta \simeq 2 \gamma. \label{2p1_crit:delta_gamma_relation}
\end{equation}
Note that from Eq.~\eqref{2p1:MAHrhomax_relation}, $c_\rho c_M = \mathcal{C}$.
The scaling laws are illustrated in Fig.~\ref{fig:2d:2p1:mass_ricci_scaling},
which shows $\ln M_\text{AH}$ and $-\ln \rho_\text{max}$ against $\ln|p-p_\star|$.

In Table~\ref{table:C_delta_2gamma}, we record the values of
$\mathcal{C}$, $\delta$, and $2 \gamma$ for different initial data and
$\tilde \mu$. We find that the relation in Eq.~\eqref{2p1:MAHrhomax_relation}
holds independently of the initial data.
On the other hand, we cannot exclude that the exponents $\delta$ and
$2 \gamma$ depend weakly on $\tilde\mu$. Similarly, we find that
$\mathcal{C}$ may weakly depend on both the initial data and $\tilde{\mu}$.

We have checked that our results are not affected by the location of
the (unphysical) outer boundary $R=R_{\rm max}$ by performing a
bisection with $R_\text{max} \simeq 10$.

\begin{table} \centering
	\begin{tabular}{p{0.45\columnwidth}>{\centering}
			p{0.15\columnwidth}>{\centering}
			p{0.15\columnwidth}>{\centering\arraybackslash}p{0.15\columnwidth}}
		\hline \hline
		Initial data ($\kappa = 0.5$) & $\mathcal{C}$ & $\delta$ & $2 \gamma$ \\
		\hline
		Off-centered, $\tilde{\mu} = 0.01$ & 0.148 & 0.0364 & 0.0371 \\
		Off-centered, $\tilde{\mu} = 0.1$  & 0.151 & 0.0413 & 0.0409 \\
		Centered,     $\tilde{\mu} = 0.1$  & 0.170 & 0.0440 & 0.0410 \\
		Ingoing,      $\tilde{\mu} = 0.1$  & 0.156 & 0.0417 & 0.0409 \\
		Off-centered, $\tilde{\mu} = 1$    & 0.161 & 0.0410 & 0.0389 \\
		Off-centered, $\tilde{\mu} = 10$   & 0.167 & 0.0427 & 0.0391 \\
		\hline \hline
	\end{tabular}
	\caption{The values of $\mathcal{C}$, $\delta$, and $2 \gamma$
		for different initial data and $\tilde \mu$, all for
		$\kappa=0.5$. These values are obtained by fitting a
		straight line to the log-log plots from super/sub15 to
		super/sub25 data points. (We only go to super/sub22 for
		$\tilde \mu = 10$).}
	\label{table:C_delta_2gamma}
\end{table}

\subsubsection{The critical solution}

In $3+1$ dimensions or higher, the critical solution exhibiting
type~II phenomena has always been found to be either continuously or
discretely self-similar, depending on the matter field under
consideration. The critical solution then only depends on $x =
R/(t_\star-t)$ (in polar-radial coordinates, with $t$
normalized to be proper time at the center) in the continuous case,
while in the discrete case, it also depends on the logarithm of
$t_\star-t$, with some period $\Delta$.

In $2+1$ dimensions, the presence of a cosmological constant is
required for black holes to exist and thus for the possibility of
critical phenomena to occur. Therefore, the Einstein equations are not
scale-free, and as a consequence the critical solution cannot be
\textit{exactly} continuously or discretely self-similar. However, as
the solution contracts to increasingly smaller scales, one expects the
effect of the cosmological constant to become dynamically irrelevant.
The critical solution could then again be approximated by an expansion
in powers of (length scale of the solution)/$\ell$, where the zeroth-order
term is a self-similar solution of the $\Lambda=0$ Einstein and
matter equations. This is actually the case for the massless scalar
field \cite{Jalmuzna15}, where the critical solution is well
approximated near the center by a continuously self-similar solution
of the $\Lambda=0$ field equations, but where the presence of
$\Lambda$ becomes relevant near the light cone.

However, it is shown in Appendix~\ref{appendix:appendixA} that a
regular self-similar solution does not exist for the perfect fluid
with a barotropic equation of state in $2+1$ (in contrast to higher
dimensions, where it is the type~II critical solution). Given
that we have type~II critical collapse nevertheless, this raises
the question of what form $s(t)$ takes.

In order to quantify $s(t)$, one can consider ``candidate'' functions $R_M(t)$
[defined previously; see Eq.~\eqref{rho0_MOB_RM}] and $R_v(t)$, defined by
\begin{equation}
v(t,R_v(t)) := 0.
\end{equation}
Both of these functions are expected to be related to the size of the
solution $s(t)$. We characterize the minimum size of the solution by
$R_{M\text{,min}} := \min\limits_t R_M(t)$. It turns out that this
also scales as suggested by its dimension, i.e.,
\begin{equation}
R_\text{M,min} = \ell c_R |p-p_\star|^\gamma,
\end{equation}
where $c_R$ is a dimensionless constant; see
Fig.~\ref{fig:2d:2p1:mass_ricci_scaling}.

\begin{figure*} 
	\includegraphics[width=1.\columnwidth, height=0.9\columnwidth]
	{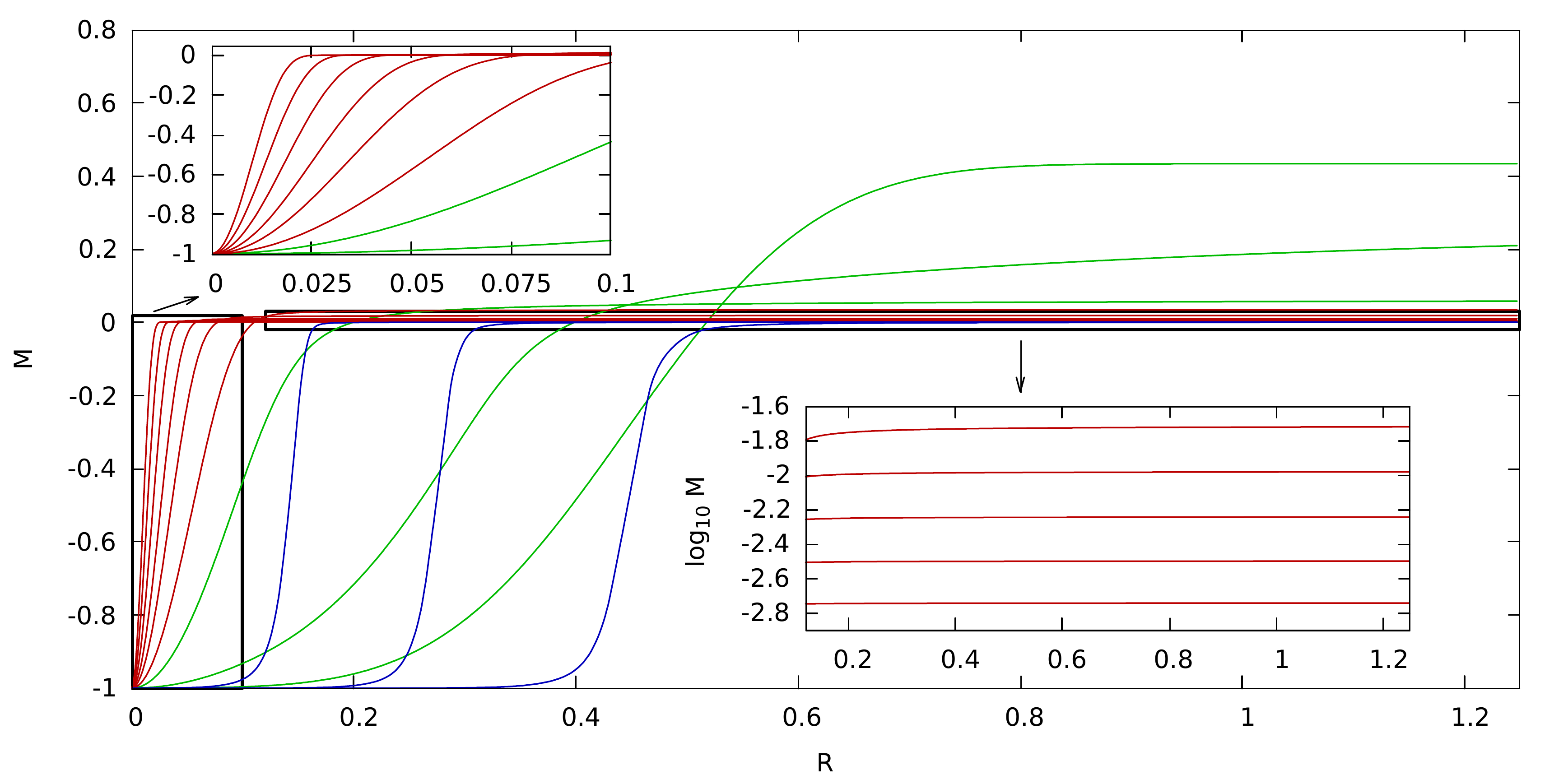}
	\includegraphics[width=1.\columnwidth, height=0.9\columnwidth]
	{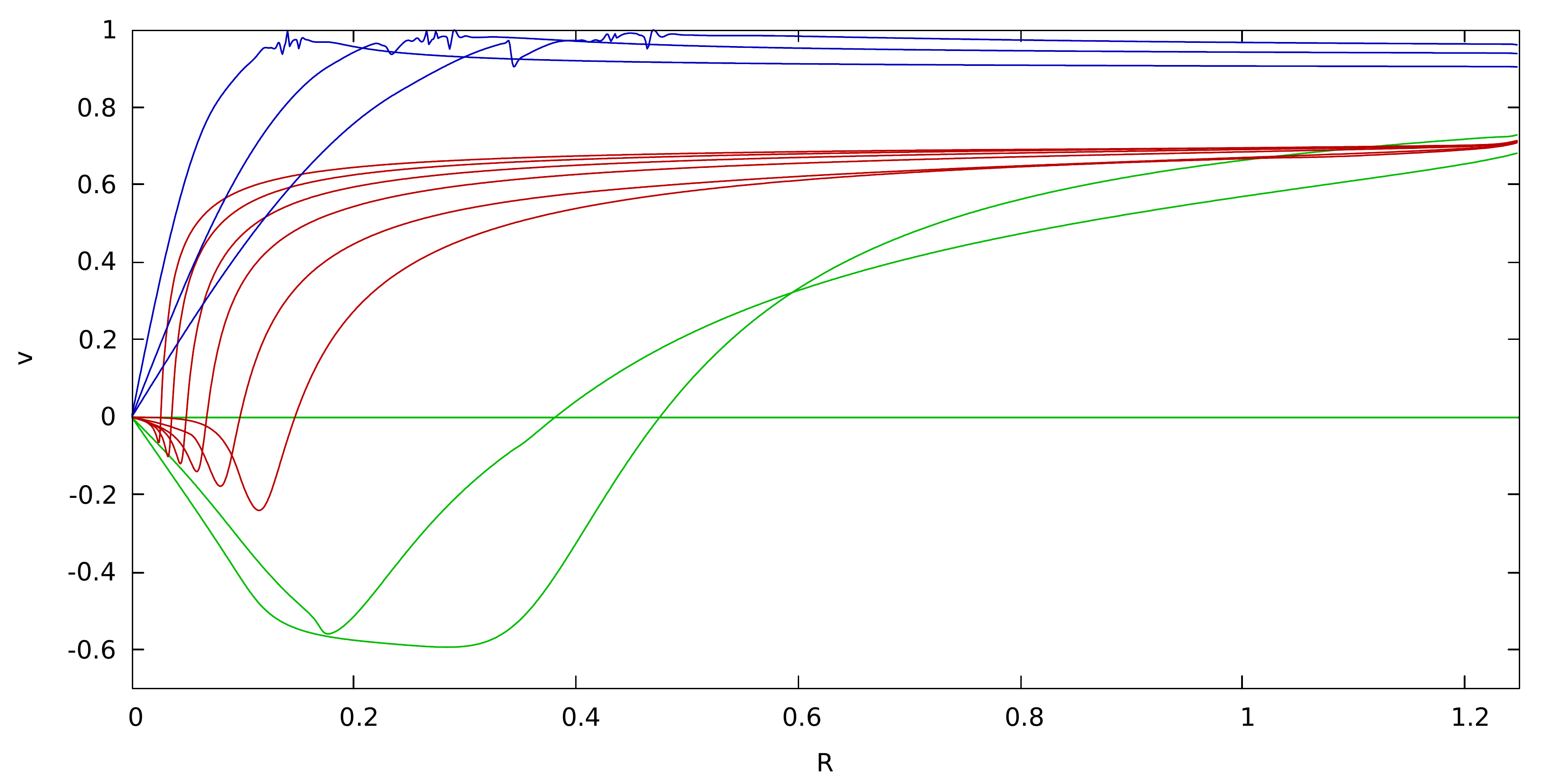}
	\caption{Plots of $M(t_i,R)$ (left) and $v(t_i,R)$ (right) at
		different times $t_i \simeq 0.$, $0.25$, $0.5$, $\cdots$, $2.25$,
		$2.37$, $2.5$ for our best subcritical data. The density
		profile shrinks and a large part of the total mass leaves
		the numerical domain (green). The evolution enters a
		critical regime (red) before dispersing (blue). In the bottom
		inset, the logarithm of the mass at large radius is plotted
		during the critical regime to show that it is almost
		constant in space and decays exponentially to zero. In the
		top inset, note that the contraction slows down. During
		dispersion, the velocity is close to $1$ at $R \simeq0.2$,
		causing numerical errors.}
	\label{fig:2d:2p1:mass_velocity}
\end{figure*}

\begin{figure*}
	\includegraphics[width=2.1\columnwidth, height=1.3\columnwidth]
	{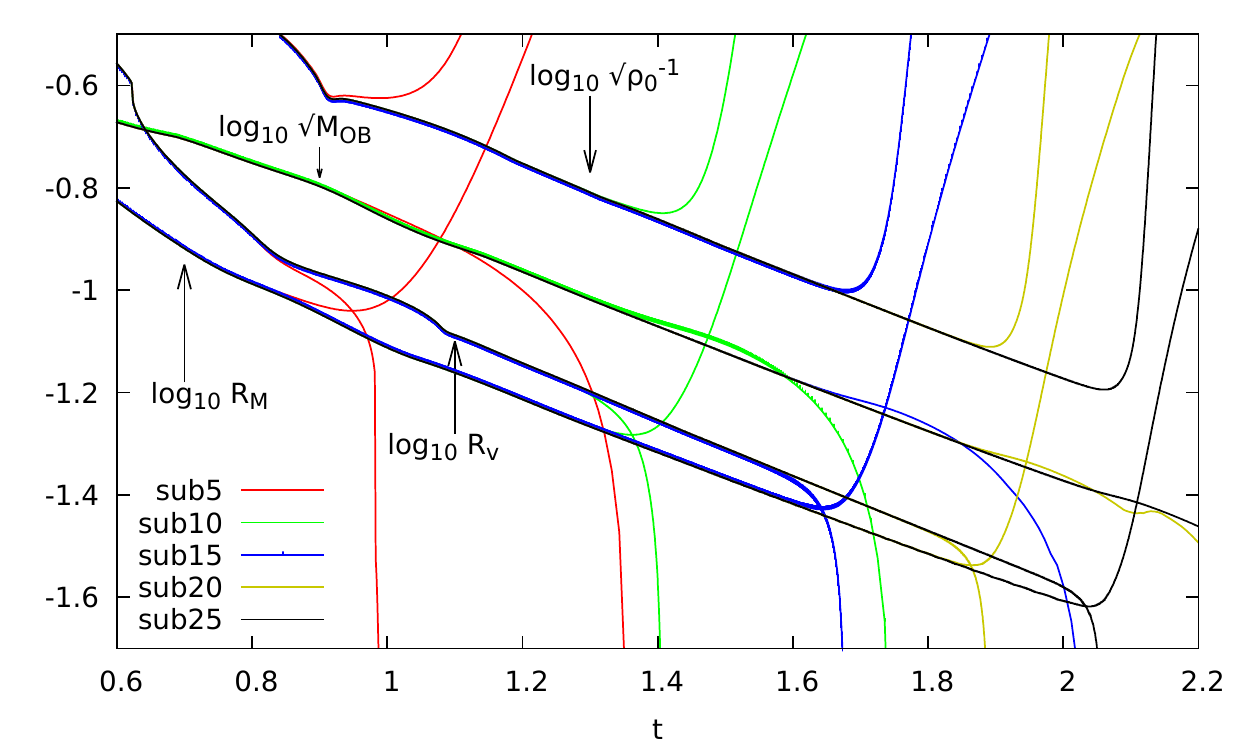}
	\caption{Log plot of $R_M(t)$, $R_v(t)$,
		$\sqrt{\rho_0^{-1}(t)}$, and $\sqrt{M_\text{OB}(t)}$, for
		sub5 to sub25 off-centered initial data. We observe that as
		we fine-tune to the black-hole threshold, the solution
		approaches an intermediate attractor solution in which
		$R_M$, $R_v$, $\rho_0^{-1}$, and $M_\text{OB}$ decrease
		exponentially. Less fine-tuned initial data peel off from
		this critical line sooner than more fine-tuned data, leading
		to critical scaling of the maximum density, etc.}
	\label{fig:RM_Rv_MOB_sub}
\end{figure*}

In $2+1$ dimensions, the total mass of the spacetime must be positive
for black holes to form. It is therefore instructive to see the
evolution of the mass in the case of near-critical data. In
Fig.~\ref{fig:2d:2p1:mass_velocity}, we plot $M(t_i,R)$ (left panel)
and $v(t_i, R)$ (right panel) for sub25 off-centered initial data at
times $t_i = 0$, $0.25$, $0.5$, $\cdots$, $2.25$, $2.37$, $2.5$. Near
the initial time, one part of the initial data shrinks, while the
other part leaves the numerical domain, causing the mass at the
numerical outer boundary to quickly decrease (green). As the data
approaches a critical regime (red), the mass profile shrinks with $t$
for $R \le R_M(t)$, while for $R \ge R_v(t)$, the mass is
approximately constant in $R$ but asymptotes to $0^+$ exponentially in
$t$ (see inset). The in-between region, $R_M(t) < R < R_v(t)$, is a
transition region whose width shrinks with $t$; see
Fig.~\ref{fig:RM_Rv_MOB_sub}. For our best subcritical data, we find
$t_\# \simeq 2.1$, after which the mass disperses (blue) and the
velocity is a positive function of $R$ and attains values close to $1$
(the speed of light).

From dimensional analysis, the quantities $R_M(t)$ and
$R_v(t)$, as well as the central proper density $\rho_0(t)$, are
expected to be related to $s(t)$ as
\begin{equation}
\rho_{0} \sim s(t)^{-2}, \qquad R_{M}(t) \sim R_v(t) \sim s(t).
\end{equation}

In Fig.~\ref{fig:RM_Rv_MOB_sub}, we plot the logarithms of $R_M(t)$,
$R_v(t)$, $\sqrt{\rho_0^{-1}(t)}$, and $\sqrt{M_\text{OB}(t)}$
for sub5, sub10, sub15, sub20 and sub25
data. We find that these quantities are exponential functions of $t$,
thus suggesting that $s(t)$ should also be an exponential. 
It should be noted here that for our best subcritical data, the
duration of the critical regime, $\Delta t \simeq 1$, is sufficiently
long to distinguish an exponential from a power law.

The exponential scaling lasts longer the more fine-tuned the initial
data is to the black-hole threshold, while less fine-tuned initial data
peel off sooner. This indicates that the critical solution has a single
growing mode that is being increasingly suppressed as we fine-tune to the
black-hole threshold.

It is useful to compare this plot with the left plot of
Fig.~\ref{fig:rho0rmmobsubkappa04}, which was obtained with the same
initial data and limiter but with $\kappa=0.4$. There the
proxies for $s(t)$ were approximately constant, while here we see
clear exponential shrinking. Note that we do not observe here any
numerical instability originating from the numerical outer boundary,
as we did for $\kappa \leq 0.42$. The reason for this is that for
$\kappa = 0.5$, near the numerical outer boundary, all three
characteristic speeds of the fluid are positive.

A striking feature implied by the fact that $s(t)$ is exponential
instead of polynomial in $t$ is that $t_\star = \infty$. In fact, by
the time the solution is entering the critical regime (at $t \simeq
1.1$), the speed of the contraction is small, with $\dot{R}_M,
\dot{R}_v \sim \mathcal{O}\bracket{10^{-2}}$, and decreasing
exponentially; see Fig.~\ref{fig:RM_Rv_MOB_sub}. In parallel, the
maximum absolute value of the velocity in $R < R_v(t)$ also quickly
decreases; see Fig.~\ref{fig:2d:2p1:mass_velocity} (right panel).

As a consequence, if the critical solution is of the form
$Z_\star(R/s(t))$, then, as $t \to t_\star=\infty$, the critical
solution is essentially static near the center, so that near $t \simeq
t_\#$, it can be expanded in powers of $\dot{s}$, and the leading-order
term is then the static solution. The critical solution is in this
sense \textit{quasistatic}.

To leading order in a formal expansion in $\dot s$ (noting that $\dot s$
is dimensionless), the quasistatic solutions are then approximated by
\begin{equation}
Z_\star(t,R) \simeq \check Z\bracket{{R \over s(t)}, -\Lambda s(t)^2},
\label{2p1:quasistaticZ}
\end{equation}
where $s(t)$ is now slowly time dependent. In particular,
near the center, $\check
Z(x,\mu)\simeq Z(x,0)$ as $\mu \to 0$, and so the family of static
solutions becomes asymptotically scale invariant as $\mu:=-\Lambda
s^2=s^2/l^2 \ll 1$, meaning that the size $s$ of the solution is
much smaller than the cosmological length scale.

Since by definition the velocity vanishes for the static solution, one
expects the velocity profile for the critical solution to be of the form
\begin{equation}
v_\star(t,R) \simeq \dot{s}(t)\, \check v_1\bracket{{R \over s(t)},
	-\Lambda s(t)^2},
\label{2p1:quasistaticv}
\end{equation}
to leading order in $\dot{s}$. In Appendix~\ref{appendix:appendixC},
we give explicit expressions for $\check Z$ and $\check v_1$.

Let us therefore model the critical solution as a quasistatic solution,
given to leading order by Eqs.~\eqref{2p1:quasistaticZ},
\eqref{2p1:quasistaticv}, and
\begin{equation}
s(t) \equiv s_0 e^{-\nu {t \over \ell}},
\label{st_ansatz}
\end{equation}
where $s_0$ has dimension length, while $\nu$ is dimensionless. These
two parameters are fixed by imposing
Eq.~\eqref{2p1:num_exact_fit_criteria} at times $t = 1.1$ and $t=1.9$,
which roughly mark the beginning and end of the critical regime. This
gives $s_0 \simeq 0.22$ and $\nu \simeq 0.78$.
We find that $\nu$ is the same for our three different families of initial
data, which gives some evidence that it is universal.

\begin{figure*}
	\includegraphics[width=1.\columnwidth,height=0.9\columnwidth]
	{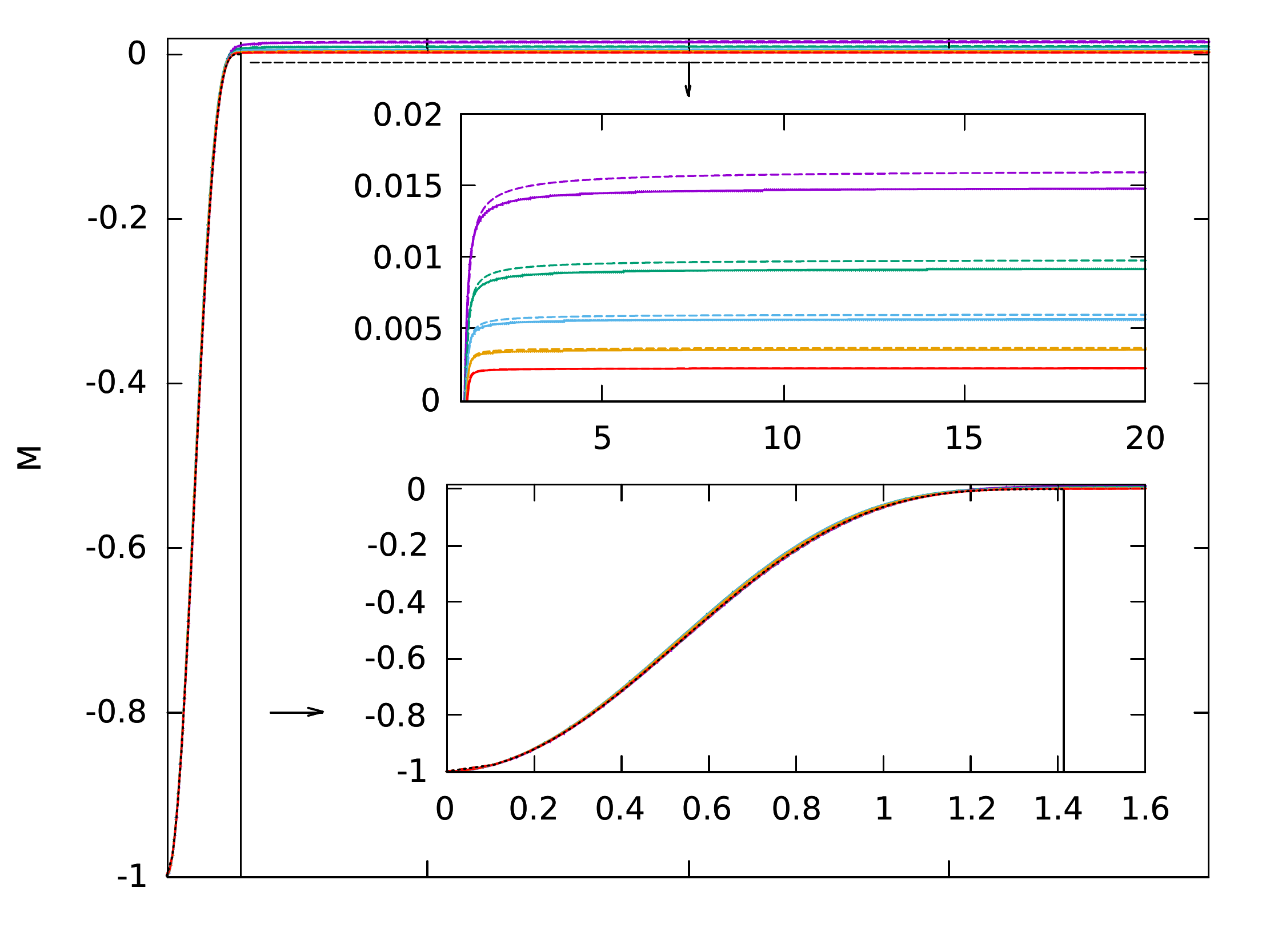}
	\includegraphics[width=1.\columnwidth,height=0.9\columnwidth]
	{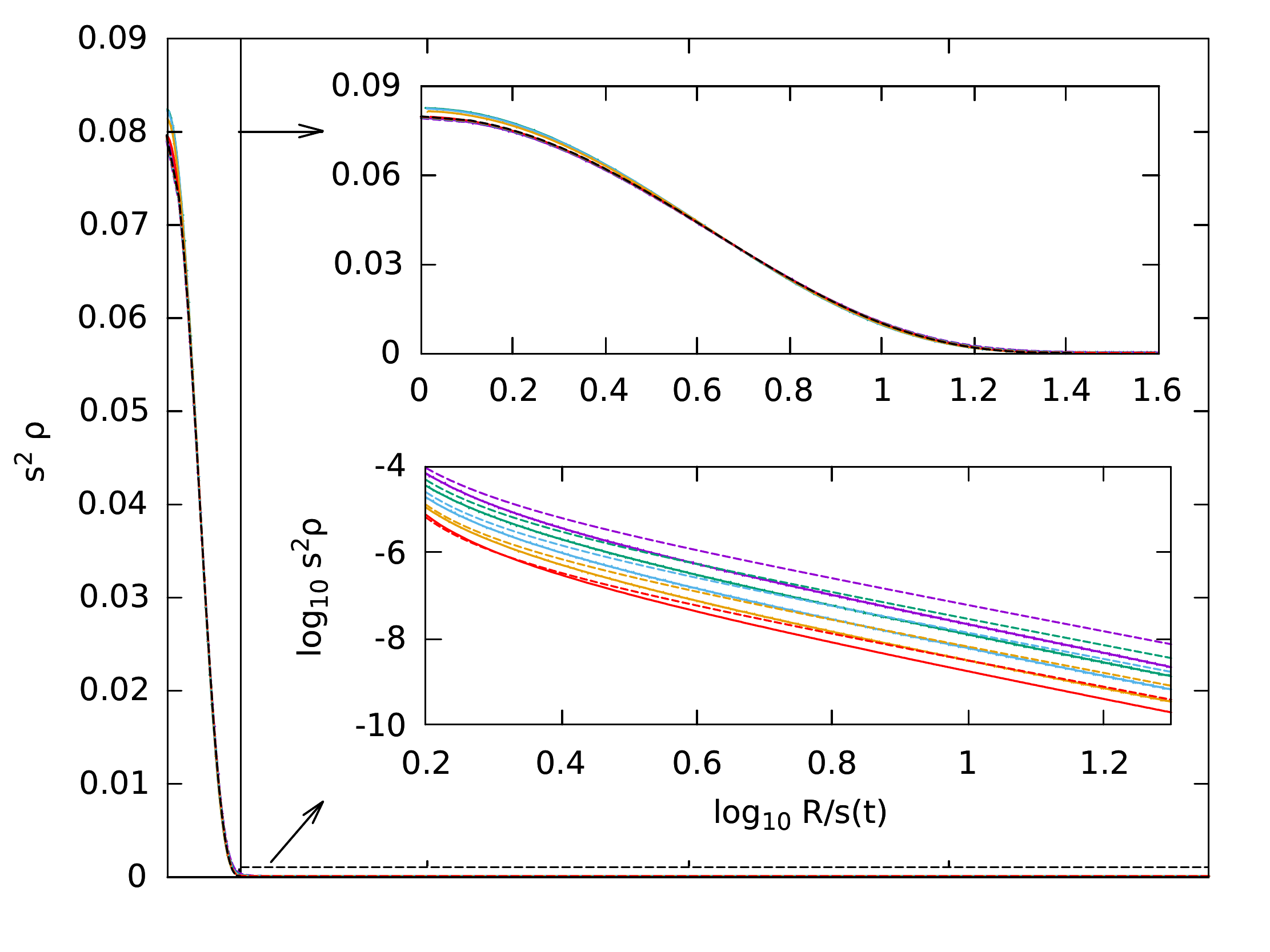}
	\includegraphics[width=1.\columnwidth,height=0.9\columnwidth]
	{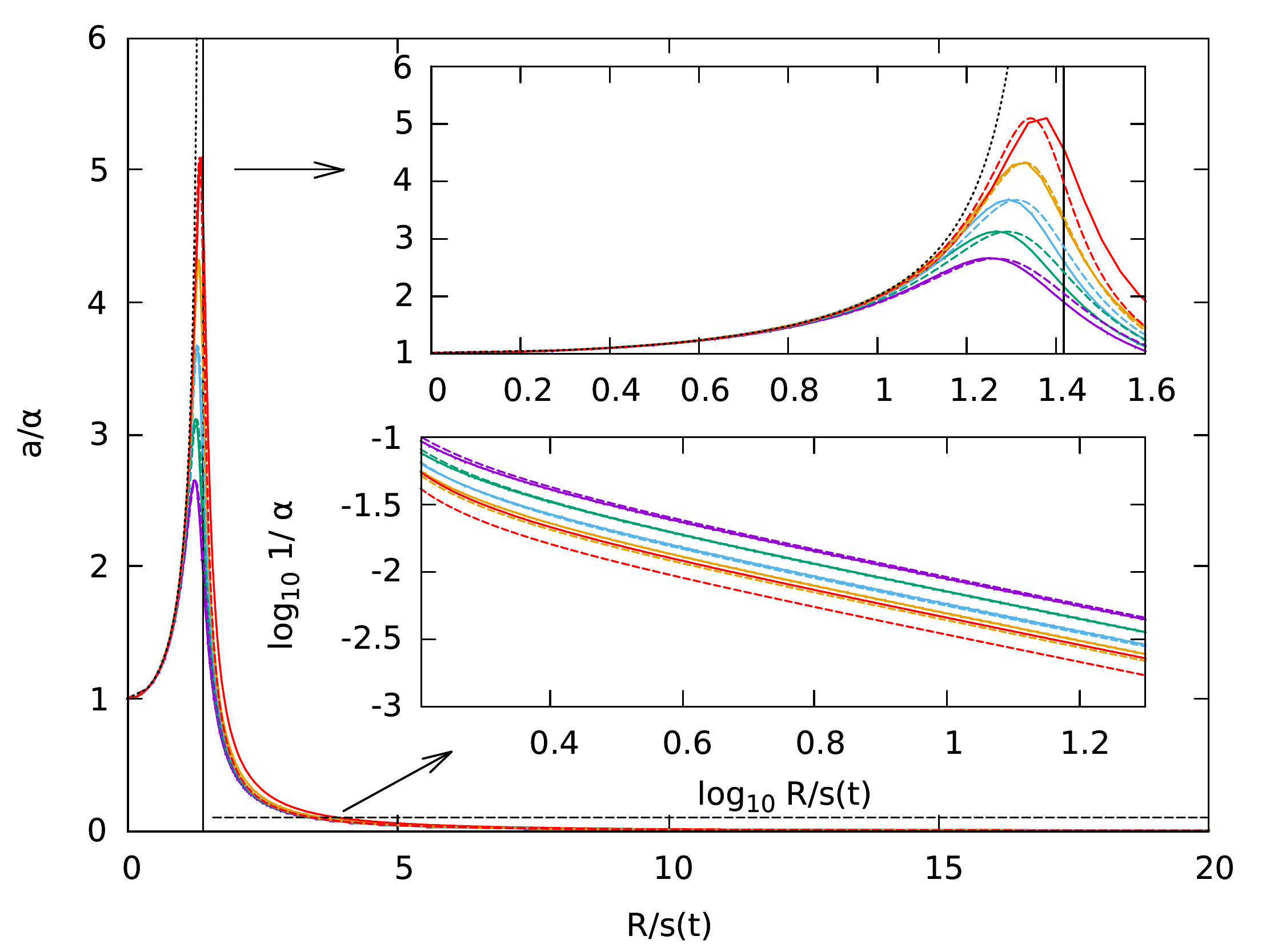}
	\includegraphics[width=1.\columnwidth,height=0.9\columnwidth]
	{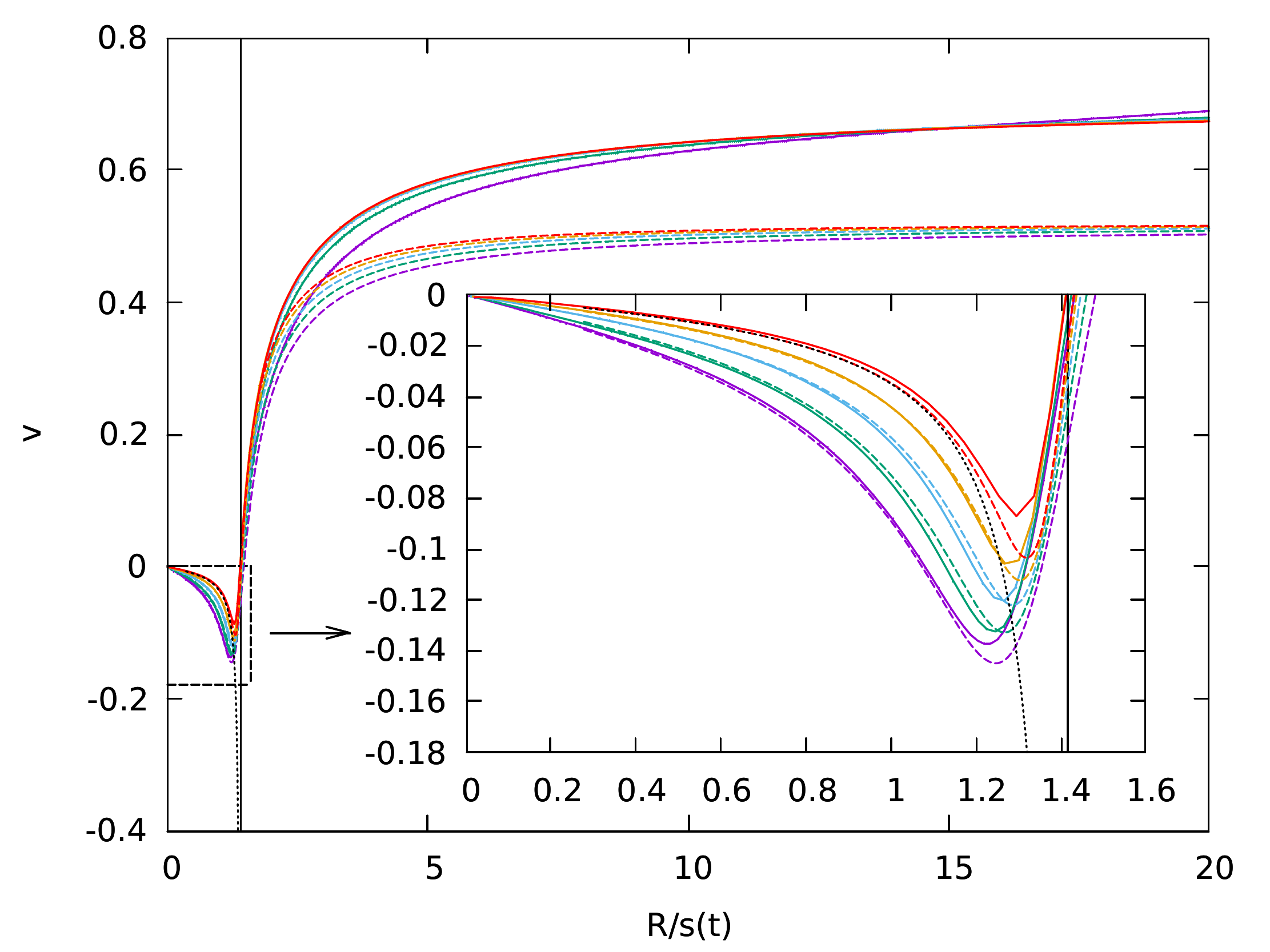}
	\caption{Numerical solution (solid colored lines) for
		$M$, $s^2 \rho$, $a/\alpha$,
		and $v$, plotted against $R/(s(t)$, at different
		times during the critical regime, $t\simeq 1.1$, $1.3$,
		$1.5$, $1.7$, and $1.9$. We have made
		a fit for $s_0$ and $\nu$ in $s(t) = s_0 e^{-\nu t/\ell}$.
		For comparison, we plot the leading-order term
		of the quasistatic $\Lambda=0$ solution (black dotted line)
		and the $\Lambda<0$ solution (colored dotted lines). The
		vertical black line corresponds to the location of the
		surface in the $\Lambda=0$ exact static solution. For
		$\Lambda<0$, there is no sharp surface.}
	\label{fig:check_scale_expo_fit}
\end{figure*}

In Fig.~\ref{fig:check_scale_expo_fit}, the critical solution is then
compared to the leading-order term of the $\Lambda=0$ (black dotted)
and $\Lambda<0$ (colored dotted) quasistatic solution in terms of $x
:= R/s(t)$.

Inside the star, the numerical solution is approximately a
function of $x$ only, implying that it is well approximated by the
$\Lambda=0$ family of static solutions. In the atmosphere, this is
not true even for small $\Lambda$, for two separate reasons.

First, the $\Lambda=0$ solution breaks down as an approximation to the
$\Lambda<0$ one at $x = x_\star$, where the $\Lambda=0$
solution has a surface, whereas the $\Lambda<0$ solutions transition
to an atmosphere. Bringing in the explicit $\Lambda$ dependence
through the second argument of $\check Z$ then also brings in a
dependence on time, as well as $x$.

Second, and more importantly, the quasistatic approximation still
holds out to the beginning of the atmosphere but we notice
that in Fig.~\ref{fig:check_scale_expo_fit}, the quasistatic
approximation systematically underestimates the falloff rate of the
density and the velocity. In other words, the true critical solution
achieves the same outgoing mass flux with a thinner atmosphere moving
more relativistically than the quasistatic approximation. This means
that a different ansatz than the quasistatic one should be made in
this regime in order to correctly model the behavior of the solution
there, and the two approximations should be matched in a
transition region, in a similar spirit as for the massless scalar
field case in Ref.~\cite{Jalmuzna15}.

Since the mass in the atmosphere is approximately constant in space,
the fluid in the atmosphere can be modeled as a test fluid on a fixed
BTZ spacetime, but not assuming that the $v$ is small. An
explicit solution under this approximation is given in 
Appendix~\ref{appendix:appendixD}.

In Fig.~\ref{fig:vs2rho_constant_flux_kappa05}, we plot, as in
Fig.~\ref{fig:check_scale_expo_fit}, the numerical and quasistatic
solutions (dotted and dashed red lines, respectively) for our best
subcritical time. In blue, we add the stationary test fluid solution
where the mass is approximately constant. We find that the latter is a
suitable ansatz for this atmosphere, as it correctly models both the
relativistic speed of the fluid and the falloff rate of the density.

\begin{figure*}
	\includegraphics[width=1.\columnwidth,height=0.9\columnwidth]
	{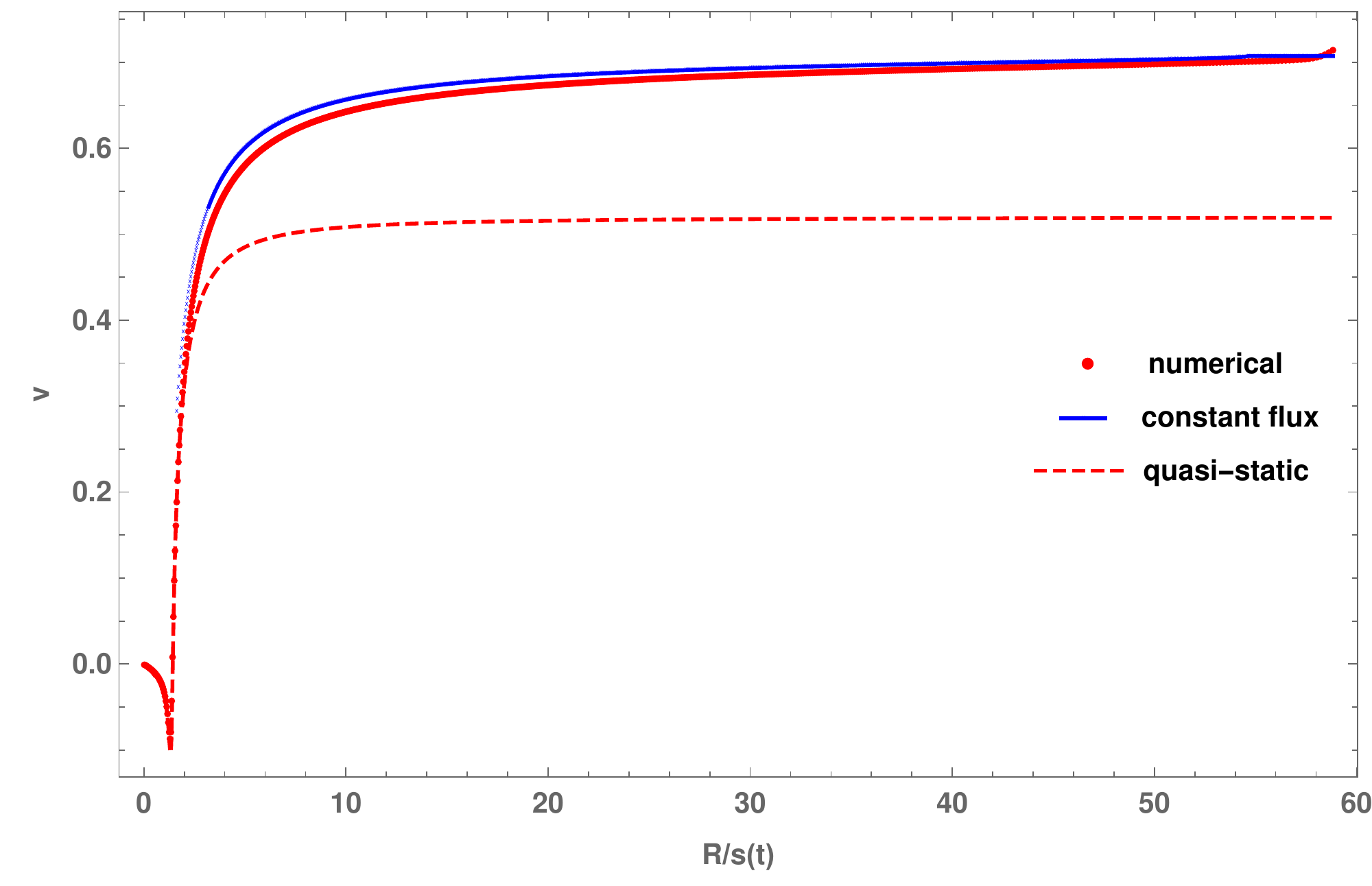}
	\includegraphics[width=1.\columnwidth,height=0.9\columnwidth]
	{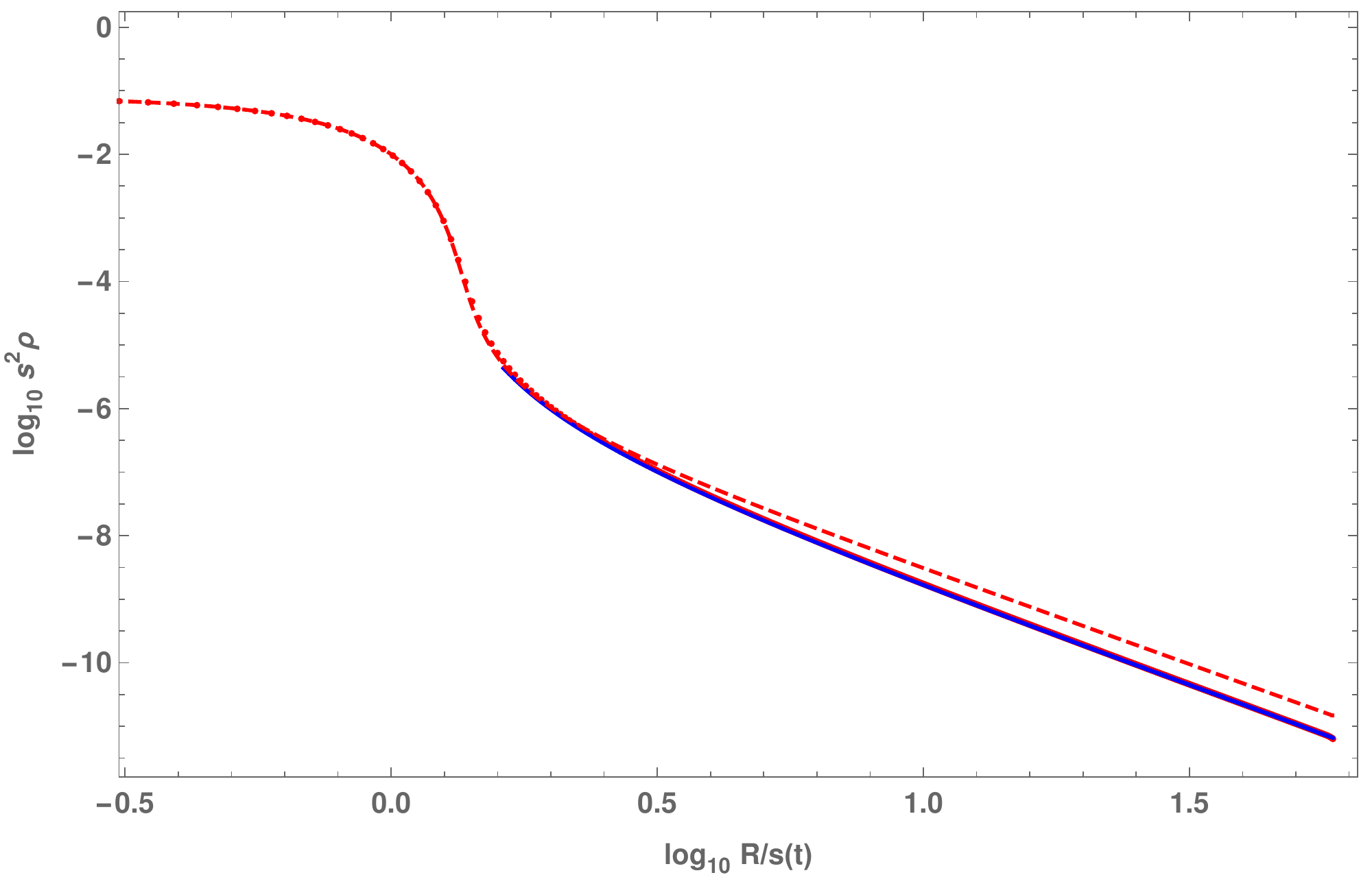}
	\caption{We compare the agreement between the numerical
		solution for our best subcritical data at $t=1.9$ (red
		line), the quasistatic solution (dashed line), and the
		test fluid solution (blue line) on the entire numerical grid.
		The quasistatic solution underestimates the asymptotic value
		of the velocity, as well as the rate of decay of the density,
		(see also Fig.~\ref{fig:check_scale_expo_fit}), while the
		constant flux solution correctly predicts these. The small upturn
		in $v$ just inside the numerical outer boundary is believed to be
		an effect of the unphysical outer boundary condition.}
	\label{fig:vs2rho_constant_flux_kappa05}
\end{figure*}

\begin{figure}
	\includegraphics[width=1.\columnwidth, height=1.\columnwidth]
	{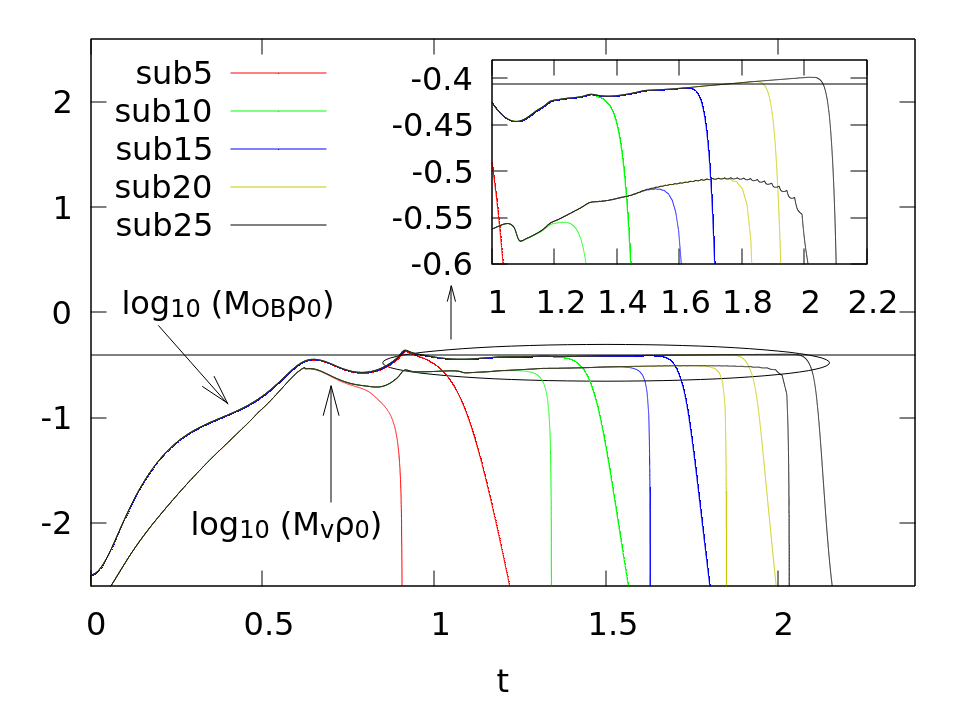}
	\caption{Log plot of $\rho_0(t) M_\text{OB}(t)$, $\rho_0(t)
		M_v(t)$ for sub5 to sub25 data. The black horizontal line
		corresponds to $\log_{10} \bracket{\check M_\infty \check \rho_0}$
		and shows that $M_\text{OB}(t)$ {is a good approximation for the
		total mass of the corresponding static solution.}}
	\label{fig:rho0_MOB_Mv}
\end{figure}

In Fig.~\ref{fig:rho0_MOB_Mv}, we plot the products
$M_\text{OB}(t)\rho_0(t)$ and $M_v(t) \rho_0(t)$, where $M_v$ is the
mass at $R_v(t)$, at different levels of fine-tuning.
The static solution has the property that in
the limit where $s \to 0$, the product $\check M_\infty \check \rho_0$
is a constant; see Eq.~\eqref{2p1:Mtotrho0}. We then expect the
product $M_\text{OB}(t) \rho_0(t)$ to approach this constant as the
solution contracts, $s(t) \to 0$, where we consider
$M_\text{OB}(t)$ as a substitute for the total mass of the static
solution. We find that during the critical regime, the product
$M_\text{OB}(t) \rho_0(t)$ is close to this theoretical value,
although in our best subcritical data it eventually becomes larger at
the end of the critical regime.

\begin{figure}
	\includegraphics[width=1.\columnwidth,height=1.\columnwidth]
	{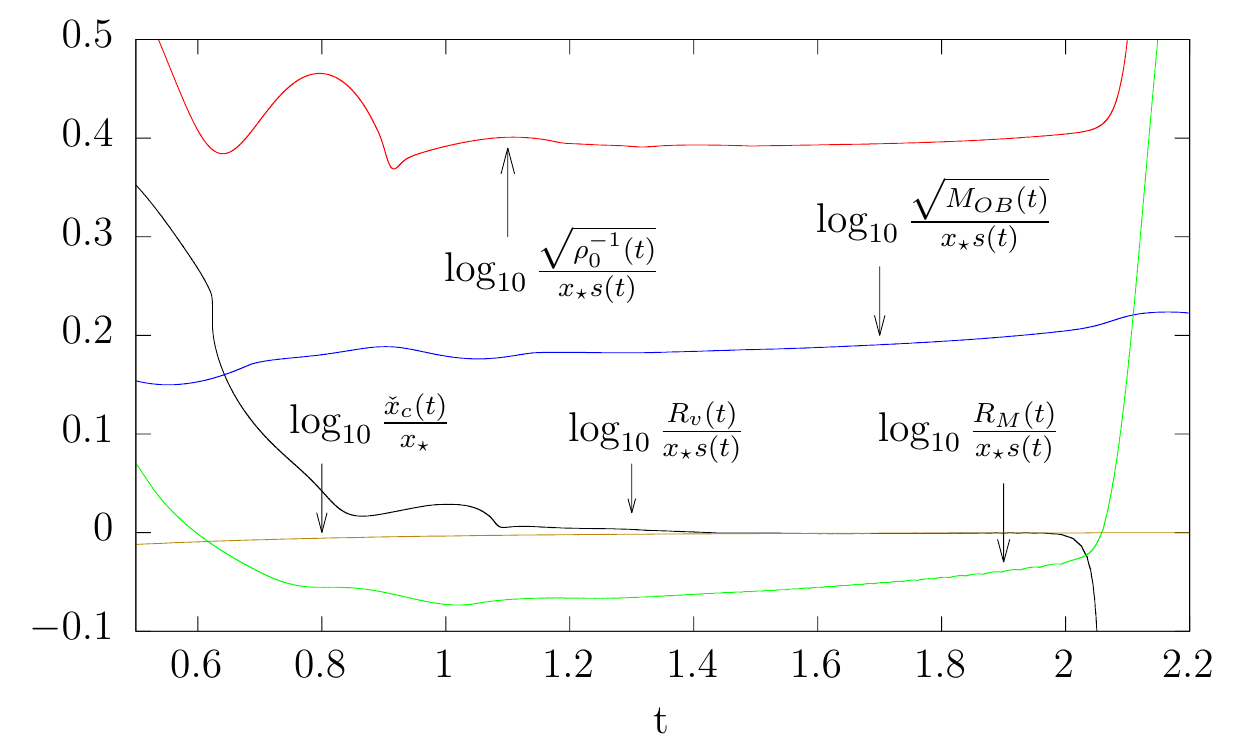}
	\caption{For sub25 data, we compare the values of $R_M(t)$,
		$R_v(t)$, $\rho_0(t)$, and $M_\text{OB}(t)$ with $x_\star
		s(t)$ and $\check x_c(t)$, where $s(t)$ is the exponential
		function fitted to the numerical data in
		Fig.~\ref{fig:check_scale_expo_fit}.}
	\label{fig:RM_Rv_sub25_fit_and_scale}
\end{figure}

In Fig.~\ref{fig:RM_Rv_sub25_fit_and_scale}, we compare the fitted
scale function $s(t)$ with the observed functions $R_v(t)$, $R_M(t)$
and $\rho_0(t)$, $M_\text{OB}(t)$. We find that $x_\star s(t)$
approximately matches $R_v(t)$ in the critical regime, although the
plot also suggests that $R_M \to x_\star s(t)$, very slowly. $s(t)$
also matches $R_M(t)$ and $\rho_0(t)M_\text{OB}(t)$, up to constant
factors. 

We also see that $R_v \simeq s(t) \check x_c$, which suggests that
$R_v \simeq R_\text{AH}$ for supercritical data.

\begin{figure*}
	\includegraphics[width=2.2\columnwidth, height=1.3\columnwidth]
	{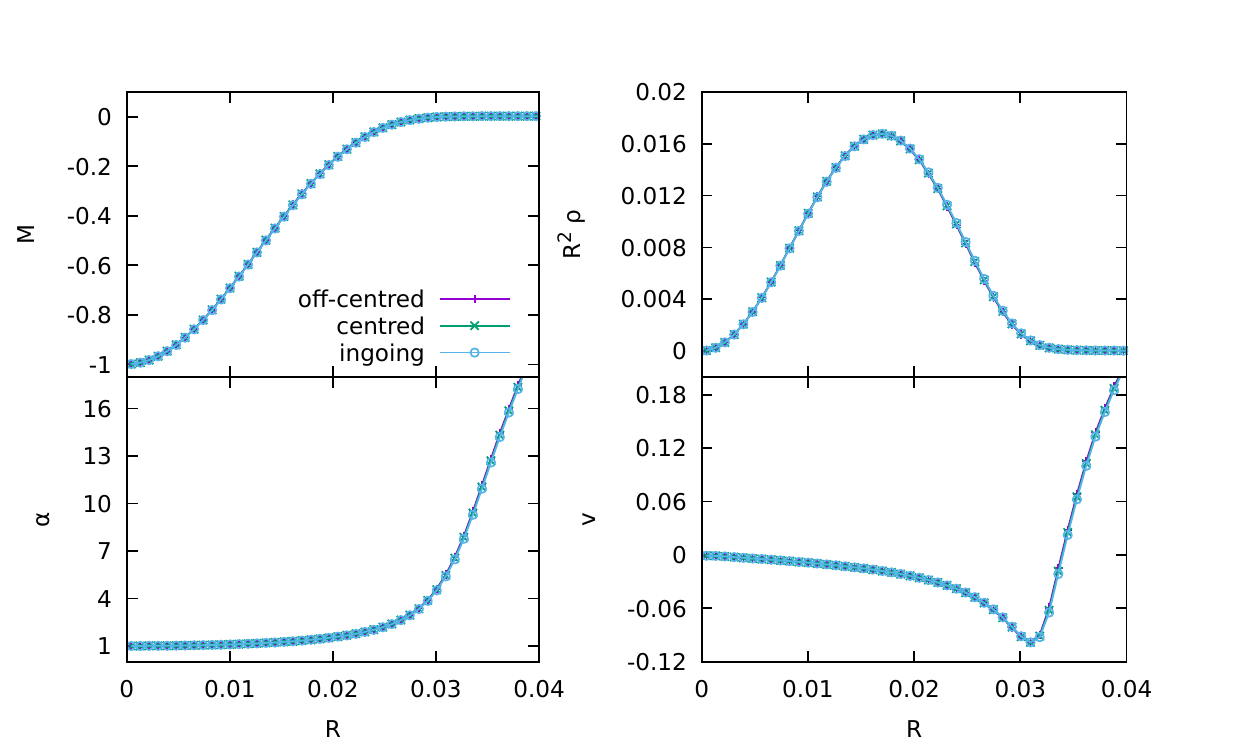}
	\caption{The profiles of $M$, $R^2 \rho$, $\alpha$, and $v$ in
		the critical intermediate attractor solution for three
		different sub25 initial data, giving some evidence for
		universality. The data have been plotted at times $t \simeq
		1.80$, $1.67$, and $1.74$, respectively, chosen to align them.}
	\label{fig:universality}
\end{figure*}

Finally, in Fig.~\ref{fig:universality}, we provide some evidence for
universality of the critical solution by plotting the numerical solution at a
fixed time for different initial data. We plot the profiles of $M$, $\alpha$,
$R^2 \rho$, and $v$ for sub25 off-centered, centered and ingoing initial
data with $\tilde{\mu}=0.1$. Since $t_\#$ depends on the
initial data, these three solutions were plotted at different times,
namely $t \simeq 1.80$, $1.67$, and $1.74$, respectively, so that the
profiles match up.

\subsubsection{Derivation of scaling laws}
\label{section:2p1_crit:mass_power_law_scaling}

In this section, we provide for a theoretical understanding of the
observed law [Eq.~\eqref{2p1:MAHrhomax_relation}].

Motivated by our numerical results, we assume that the critical
solution can, to leading order, be modeled as a quasistatic
solution, $\check Z$; see Appendix~\ref{appendix:appendixC}.

For this discussion, two properties of the family of static solution
are of importance: First, the central density scales like $\check
\rho_0 \sim \ell^{-2} \sim s^{-2}$; see Eq.~\eqref{rhobar}. Second,
the total mass of the system, assuming small $\mu \ll 1$, scales like
$\check M_\infty \sim \mu \sim s^{2}$; see Eq.~\eqref{2p1:static:Mtot}.

Since the maximum of the curvature is attained at the center, we can approximate
\begin{equation}
\rho_\text{max} \simeq \rho_{0,\text{evolved}}(t_\#) \simeq
\check \rho_{0}(-\Lambda s(t_\#)^2).
\end{equation}

On the other hand, unlike for the density, one cannot simply make the
approximation $M_\text{AH} \simeq M_\infty \simeq \check M_\infty$,
since the total mass of the system is time independent and therefore
cannot scale.

However, our numerical outer boundary does allow mass to
escape. Moreover, we have seen that in the region from the surface of
the shrinking star to the outer boundary, $M(t,r)$ is approximately
constant in $r$ and decreasing adiabatically in time, and that
this observation does not depend on the location of the numerical
outer boundary. We therefore conjecture that this atmosphere of
constant mass flux is physical.

Somewhere further out, and beyond our numerical outer boundary, we would
of course find enough mass to bring $M_\infty$ to its time-independent value.

In the intermediate regime, between $R_v(t)$ (the point beyond which
$M$ is approximately constant in space) and the outer boundary
$M_\text{OB}$, we can then approximate
\begin{equation}
M_\text{intm.}(t,R) \simeq \check M_\infty \left(-\Lambda s(t)^2\right)
\sim s(t)^2.
\end{equation}
As the black hole must form with $M>0$, and $M>0$ holds only in the
intermediate regime, not inside the star, it follows that $R_\text{AH}
> R_M(t)$. Furthermore, recall that from
Fig.~\ref{fig:RM_Rv_sub25_fit_and_scale}, in the critical regime
$\check x_c(t) \simeq R_v(t)$, implying that $R_v(t_\#) \simeq R_\text{AH}$.
It is then natural to assume that $M_\text{AH}$ takes the above value,
evaluated at $t_\#$. That is,
\begin{equation}
M_\text{AH} \simeq M_\text{intm.}(t_\#,R_\text{AH})\sim s(t_\#)^{-2}.
\end{equation}

Taking both approximations for $\rho_\text{max}$ and $M_\text{AH}$
in terms of $\check \rho_0$ and $\check M_\infty$ together into
Eq.~\eqref{2p1:Mtotrho0}, we then infer the relation in
Eq.~\eqref{2p1:MAHrhomax_relation}.

This analysis also allows us to predict that
$\mathcal{C} = (4 \pi (1-\kappa))^{-1}$. For $\kappa=0.5$, this gives
$\mathcal{C} \simeq 0.16$, which is consistent
with the numerical result; see Table~\ref{table:C_delta_2gamma}.

\subsection{Type~I-II transition}

\begin{figure*}
	\includegraphics[width=2.\columnwidth, height=1.3\columnwidth]
	{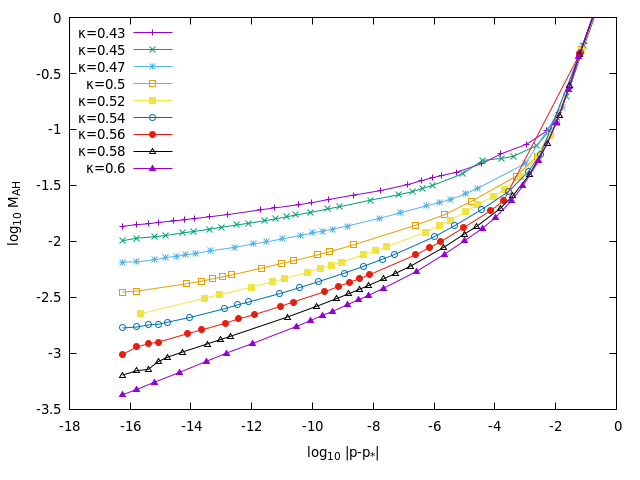}
	\caption{Apparent horizon mass scaling for different values of $\kappa
		\geq 0.43$. We find typical type~II scaling. In all cases, the
		relation $\delta=2 \gamma$ is verified. Compare the
        equivalent (but flat) scaling laws for $\kappa\leq
        0.42$ in Fig.~\ref{fig:Mscale_typeI}.}
	\label{fig:Mscale_typeII}
\end{figure*}

For arbitrarily good fine-tuning, the type~II apparent horizon
mass becomes vanishingly small, while the type~I mass is a
family-dependent constant. In the region between $\kappa=0.43$ and
$\kappa=0.5$, type~II phenomena are still observed (see
Fig.~\ref{fig:Mscale_typeII}), but this is already a transition from
type~I to type~II.

\begin{figure*}
	\includegraphics[width=1.\columnwidth]
	{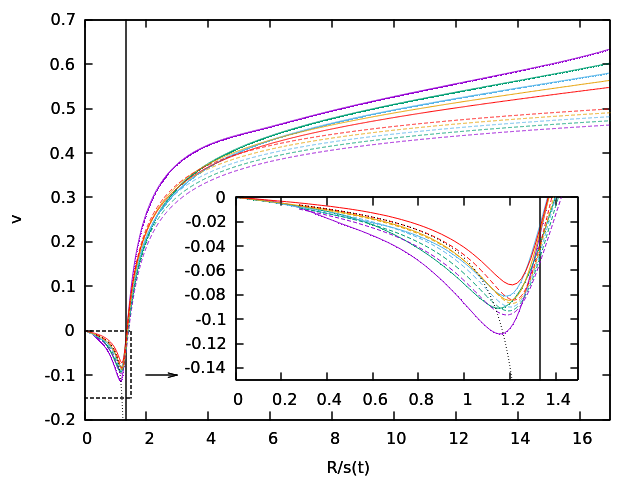}
	\includegraphics[width=1.\columnwidth]
	{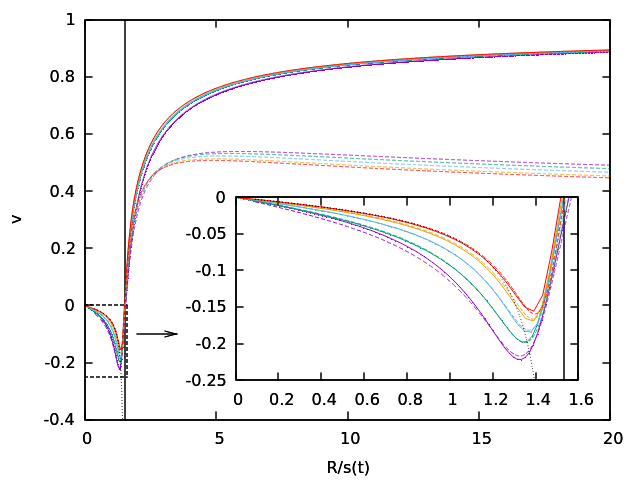}
	\caption{Comparison of the analytical and numerical expressions of
		$\check v_1$ for $\kappa=0.47$ (left) and $\kappa=0.54$ (right).
		The colored (dotted) lines follow a similar convention to
		Fig.~\ref{fig:check_scale_expo_fit}.}
	\label{fig:v1_different_kappa}
\end{figure*}

In Fig.~\ref{fig:v1_different_kappa}, we compare the agreement between
the quasistatic solution and the numerical results for $\check v_1$ for
$\kappa = 0.47$ and $0.54$, plotted at times $t = 1.3$, $1.45$, $1.6$,
$1.75$, $1.9$ and $t=0.9$, $1.0$, $1.1$, $1.2$, $1.25$, respectively.

\begin{figure*}
	\includegraphics[width=1.\columnwidth,height=0.9\columnwidth]
	{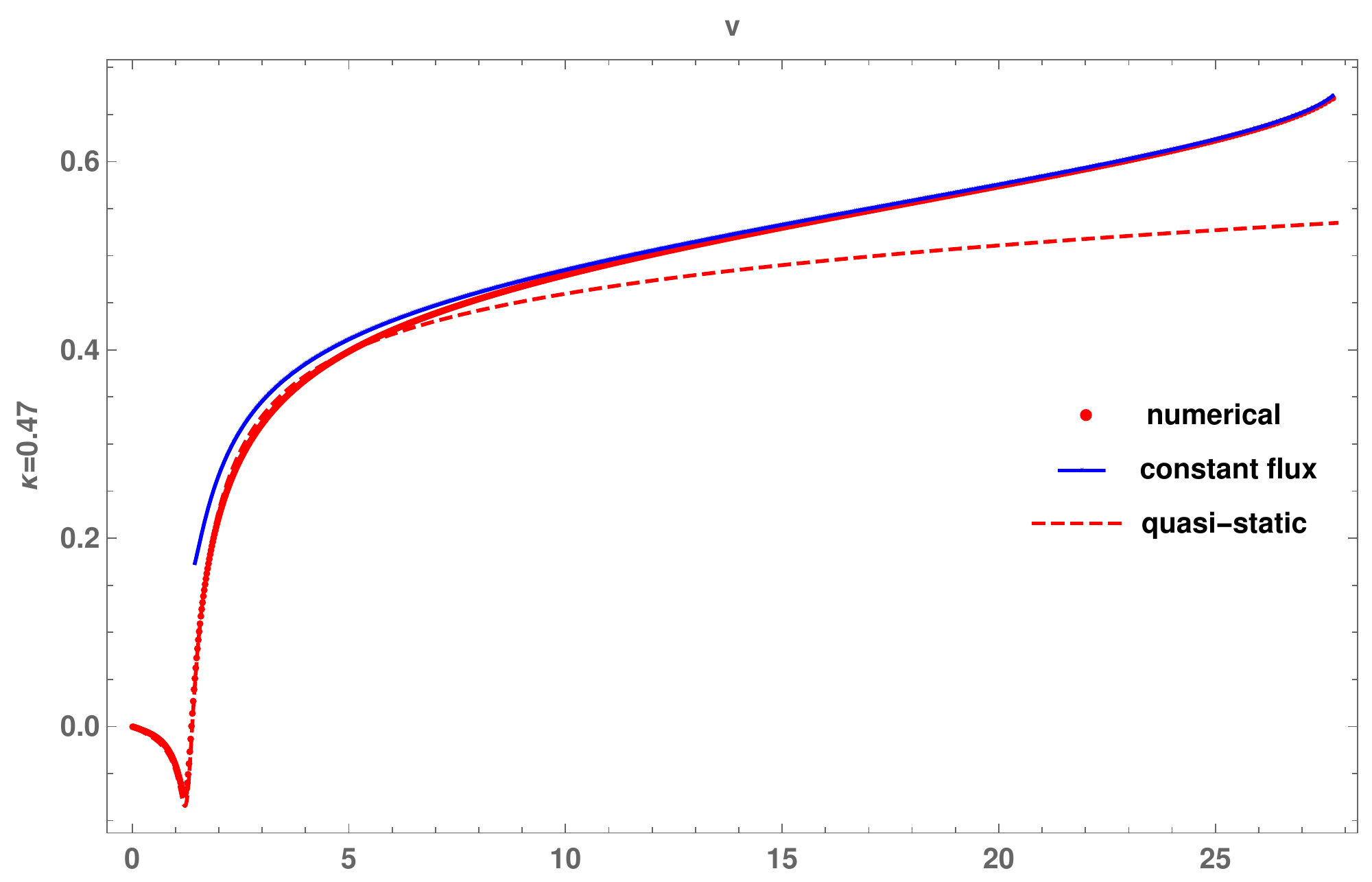}
	\includegraphics[width=1.\columnwidth,height=0.9\columnwidth]
	{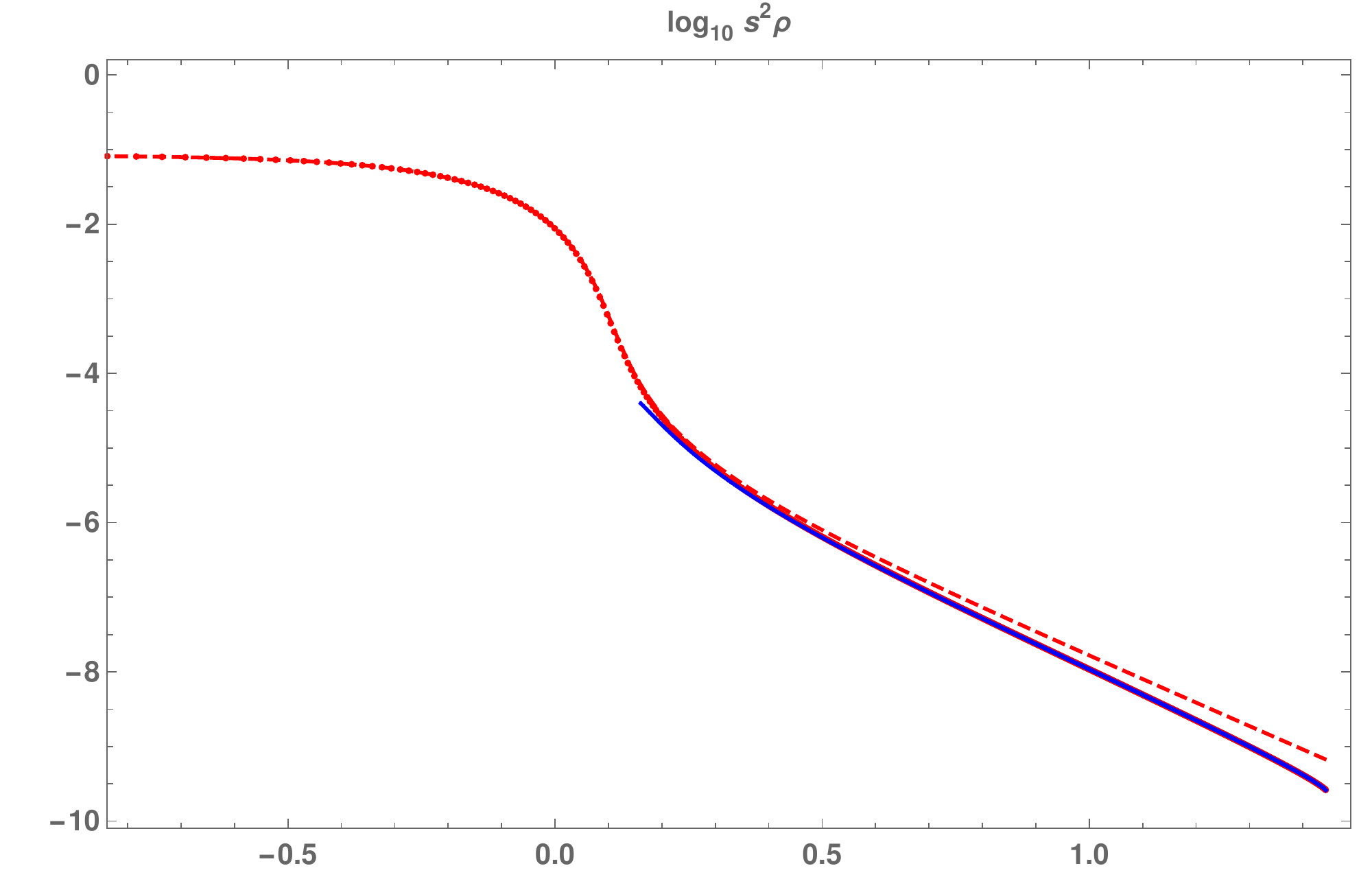}
	\includegraphics[width=1.\columnwidth,height=0.9\columnwidth]
	{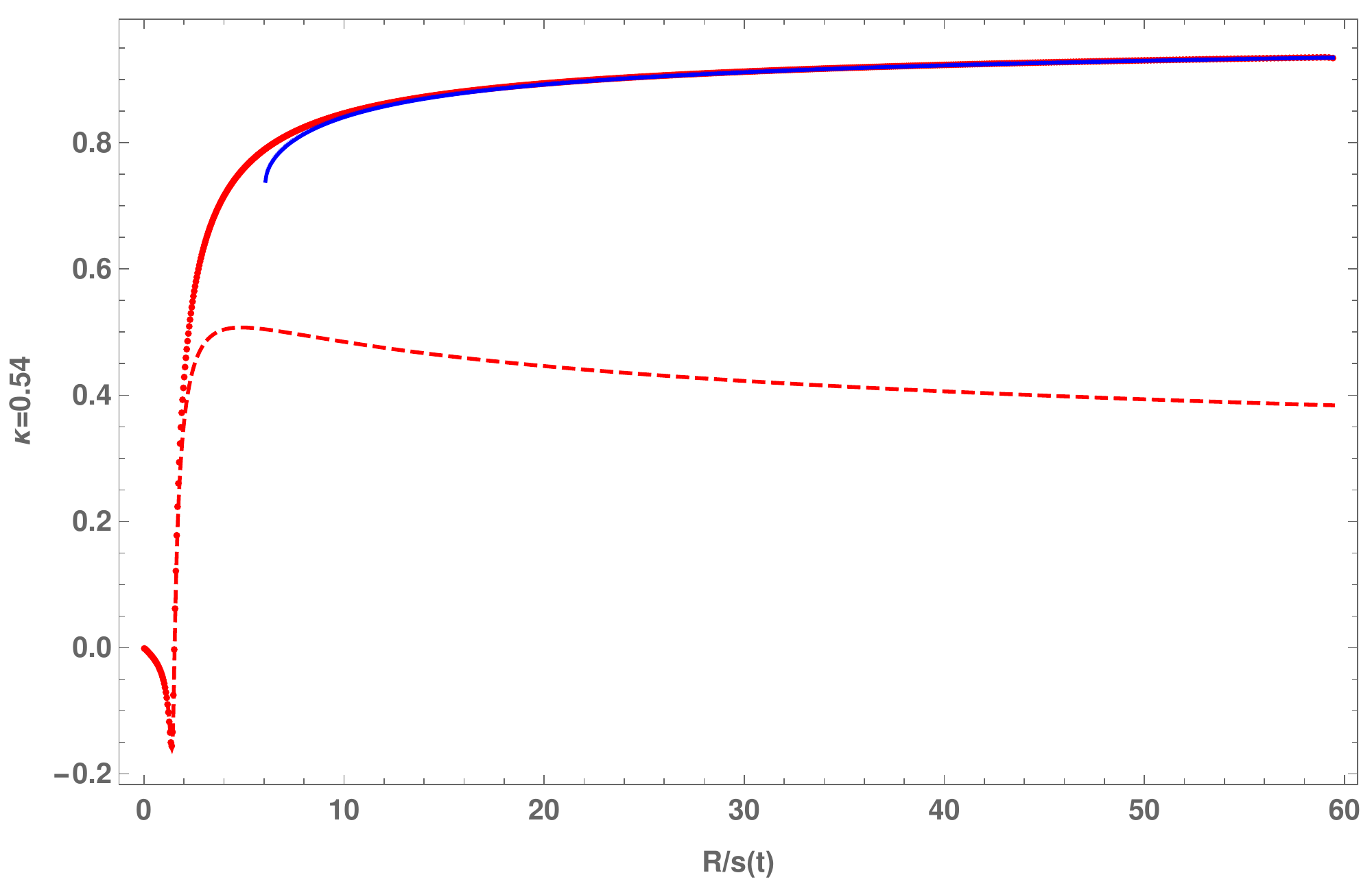}
	\includegraphics[width=1.\columnwidth,height=0.9\columnwidth]
	{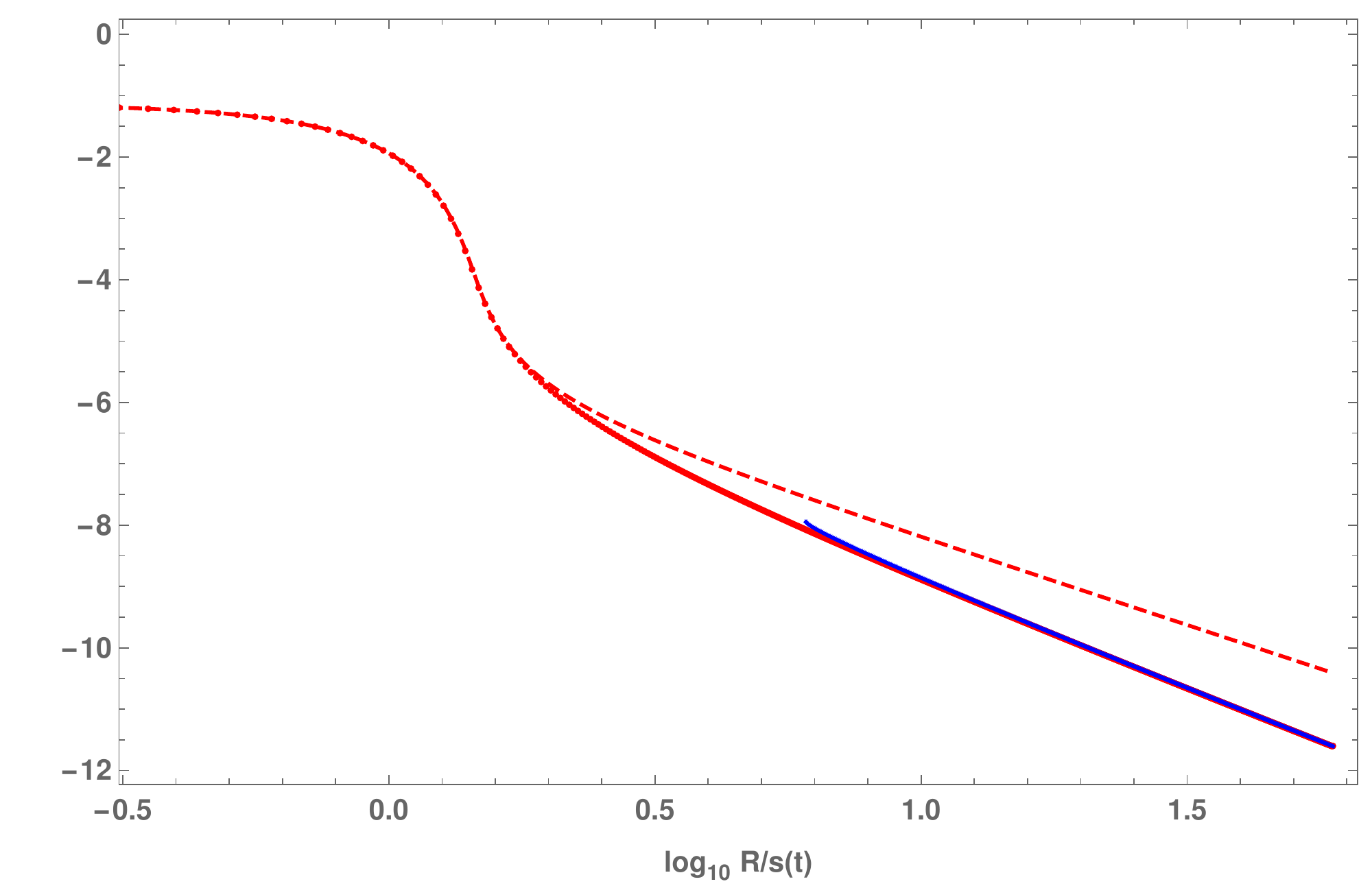}
	\caption{We compare the agreement between the numerical
		solution (red line), the quasistatic solution (dashed line), and
		the test fluid solution (blue line) for $\kappa=0.47$ (top two
		plots) and $\kappa = 0.54$ (bottom two plots) for our best
		subcritical data at $t = 1.9$ and $1.25$, respectively.
		Otherwise, as in Fig.~\ref{fig:vs2rho_constant_flux_kappa05}.}
	\label{fig:vs2rho_constant_flux_kappa047_and_054}
\end{figure*}

For $\kappa=0.54$, we find good agreement between the
numerical time evolution and the quasistatic approximation inside
the star. For $\kappa=0.47$, we find much poorer agreement, even
near the center. As for the $\kappa=0.5$ case,
the test fluid solution is a much better model for the atmosphere of
the critical solution for both {$\kappa=0.54$ and $0.47$}; see
Fig.~\ref{fig:vs2rho_constant_flux_kappa047_and_054}.

\begin{figure*}
	\includegraphics[height=1.1\columnwidth, width=0.9\linewidth]
	{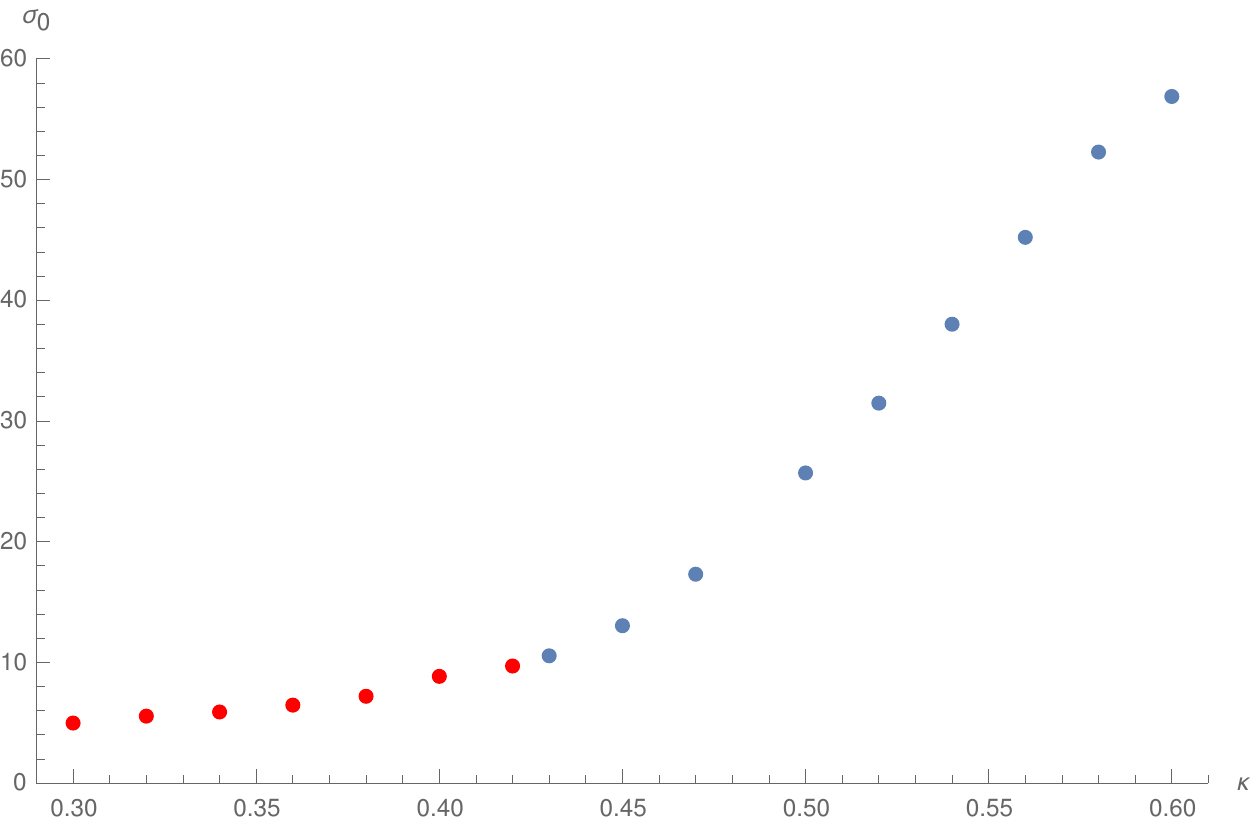}
	\includegraphics[height=0.9\columnwidth, width=0.49\linewidth]
	{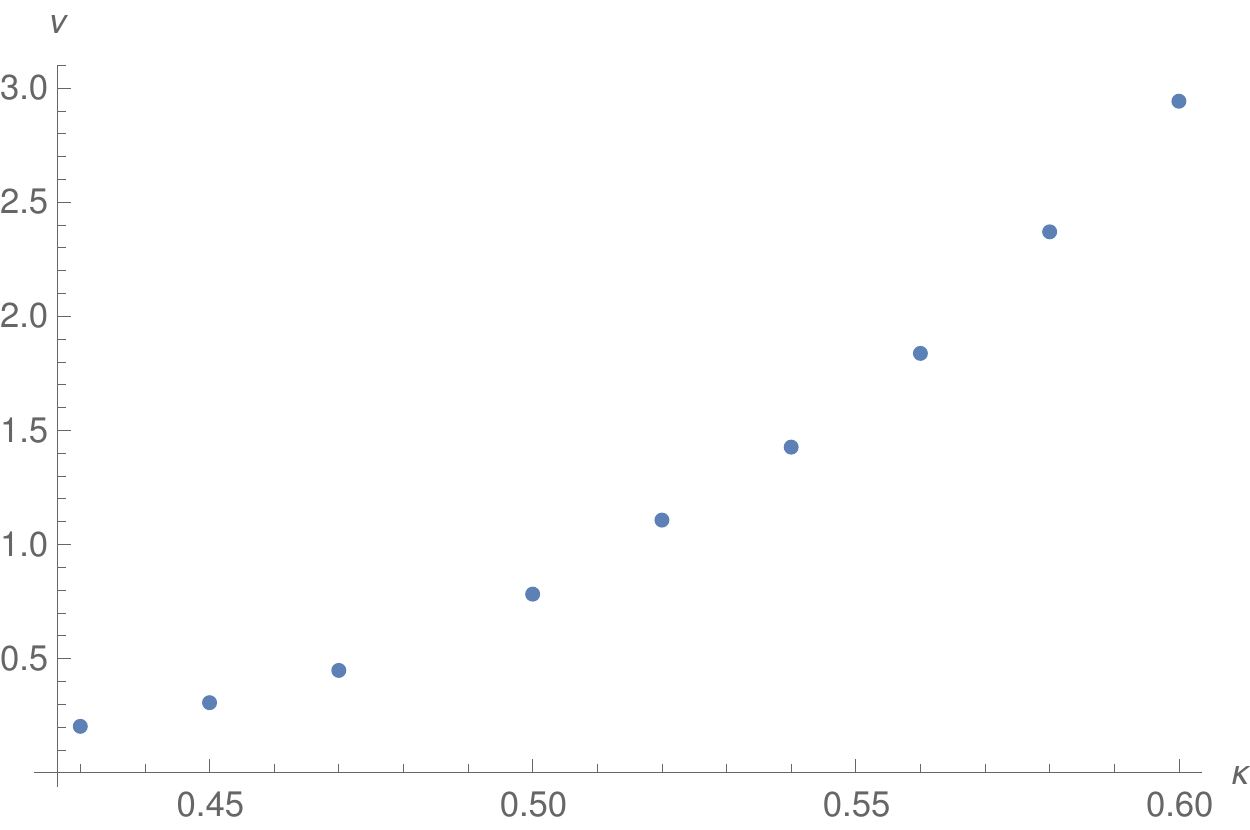}
	\includegraphics[height=0.9\columnwidth, width=0.49\linewidth]
	{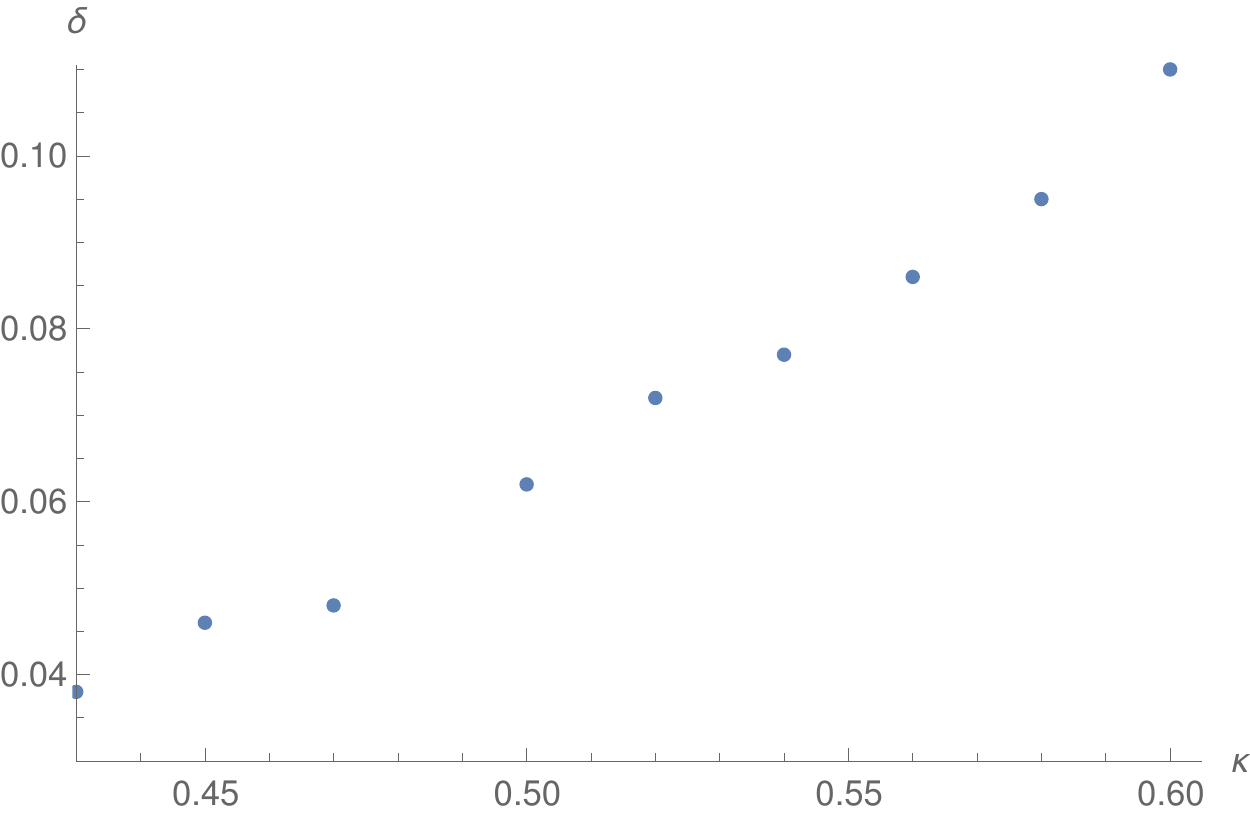}
	\caption{Plot of $\sigma_0$ (top), $\nu$ (bottom left) and
		$\delta$ (bottom right) against $\kappa$. We note that
		$\sigma_0$ appears to be continuous across the type~I (red)
		to type~II (blue) transition.}
	\label{fig:sigma0_kappa}
\end{figure*}

The exponential time dependence of the growing mode holds for both type~I
phenomena, where $s$ is constant, and type~II phenomena, where $s(t)$ is
itself exponential. In the case of type~II, 
\begin{equation}
\text{growing mode} \sim e^{\sigma_0 t\over\ell} \sim s(t)^{-\lambda_0} \sim
\left(e^{-{\nu t\over\ell}}\right)^{-\lambda_0},
\end{equation}
and so we can express $\sigma_0$ in terms of $\lambda_0$
[Eq.~\eqref{crit_lin_pert}] (or $\gamma$) and $\nu$ [Eq.~\eqref{st_ansatz}] as
\begin{equation}
\label{checkidentity}
\sigma_0 = \nu \lambda_0,
\end{equation}
and hence, with $\delta=2/\lambda_0$,
\begin{equation}
\label{deltanurelation}
\delta = {2\nu\over \sigma_0}.
\end{equation}

In Table~\ref{table:nu_delta_sigma0}, we give $\nu$, $\delta$, and
$\sigma_0$ for different values of $\kappa$, and
Eq.~\eqref{deltanurelation} is explicitly verified. The exponent $\sigma_0$
for the exponentially shrinking critical solution is computed in the
same way as in the type~I case. 

Given that $\sigma_0$ is defined both in the type~I and type~II regimes
of $\kappa$, one may ask if it is a smooth or at least
continuous function of $\kappa$ across both regimes. In
Fig.~\ref{fig:sigma0_kappa}, we plot $\sigma_0$ (top), $\nu$ (bottom
left), and $\delta$ (bottom right) against $\kappa$. The data points
are given in Tables~\ref{table:nu_sigma0_typeI} and
\ref{table:nu_delta_sigma0}. We find that $\nu$ and $\delta$ are
monotonically increasing functions of $\kappa$.  For $0.3 \leq \kappa
\leq 0.42$ and $0.5 \leq \kappa \leq 0.6$, $\sigma_0$ depends
linearly on $\kappa$. Due to shocks occurring for $\kappa \leq 0.42$
and causing a systematic error in the evaluation of $\sigma_0$ when we
use a second-order limiter, the bisections and evolutions were
performed using the Godunov limiter. We find that $\sigma_0$ is at
least continuous in the transition from type~I to type~II phenomena.

Our plots are compatible with $\delta$ vanishing at
$\kappa\simeq 0.42$ because $\nu$ vanishes there, while $\sigma_0$
remains finite. In other words, the unstable mode grows
exponentially in time in the type~I and type~II critical solution,
but this gives rise to type~II mass and curvature power-law scaling
only when the critical solution shrinks, also exponentially in
time. At the transition from type~II to type~I in the equation-of-state
parameter $\kappa$, the critical solution simply stops shrinking as
$\nu(\kappa)\to 0$.

\begin{table}
	\begin{tabular}{>{\centering}p{0.15\columnwidth}>{\centering}
		p{0.2\columnwidth}>{\centering}p{0.2\columnwidth}>{\centering}
		p{0.2\columnwidth}>{\centering\arraybackslash}p{0.15\columnwidth}}	
		\hline \hline
		$\kappa$ & $\nu$ & $\delta$ & $\sigma_0$ & 
		$\frac{\delta \sigma_0}{2 \nu}$ \\
		\hline
		$0.43$ & 0.203 & 0.038 & 10.54 & 0.99 \\
		$0.45$ & 0.307 & 0.046 & 13.03 & 0.98 \\
		$0.47$ & 0.448 & 0.048 & 17.30 & 0.93 \\
		$0.50$ & 0.782 & 0.062 & 25.69 & 1.02 \\
		$0.52$ & 1.107 & 0.072 & 31.47 & 1.02 \\
		$0.54$ & 1.427 & 0.077 & 38.01 & 1.03 \\
		$0.56$ & 1.838 & 0.086 & 45.21 & 1.06 \\
		$0.58$ & 2.370 & 0.095 & 52.27 & 1.05 \\
		$0.60$ & 2.943 & 0.110 & 56.88 & 1.06 \\
		\hline \hline
	\end{tabular}
	\caption{The values of $\nu$, $\delta$, and $\sigma_0$ as functions of
		$\kappa$. In type~II, $\nu$ is obtained directly from the critical
		solution observed in our closest-to-critical time evolutions. As in
		type~I, we have obtained $\sigma_0$ from the lifetime scaling
		[Eq.~\eqref{lifetimescaling}] of the critical solution.}
	\label{table:nu_delta_sigma0}
\end{table}

\section{Conclusions}
\label{section:conclusions}

Critical collapse in 2+1 dimensions is an intriguing toy model
for 3+1 dimensions, as in 2+1 dimensions axisymmetric solutions
depend only on radius and time, making the simulation of rotating
collapse as cheap as that of nonrotating collapse. By contrast, one
expects complications from the fact that the existence and formation
of black holes in 2+1 requires $\Lambda<0$, which breaks the
scale-invariance necessary for type~II critical phenomena.

In our time evolutions of one-parameter families of initial data,
for $\kappa\lesssim 0.42$ we find type~I critical collapse: the
maximum curvature and apparent horizon mass are constant beyond a
certain level of fine-tuning. The critical solution is static,
and the time for which it is observed scales as the logarithm of
distance to the threshold of collapse. 

For $\kappa\gtrsim 0.43$, we find type~II critical collapse: At the
threshold of (prompt) collapse, the maximum curvature diverges and the
apparent horizon mass goes to zero. However, in contrast to 3+1 and
higher dimensions, and even to scalar field collapse in 2+1, we find
that the corresponding critical solution near the center is not
self-similar, but quasistatic, moving through a family of regular
static solutions of finite mass with scale parameter $s(t)$.
Outside the central slowly shrinking star, the critical solution
is better approximated as a test fluid in a background BTZ spacetime
with $M\gtrsim 0$, moving mass away from the shrinking star at
relativistic speeds.

The only other case of a critical solution at the boundary between blowup
and dispersion that is quasistatic known to us is for a spherically symmetric
ansatz for the Yang-Mills equations in 4+1 dimensions in flat spacetime
(the critical dimension for that system, as 2+1 is for gravity). However,
we have not been able to derive the observed exponential form of $s(t)$ from
first principles along the lines of Refs.~\cite{Bizon04,Rodnianski10}.

However, there are also interesting parallels, not yet sufficiently
understood, with the self-similar critical solution in 2+1 spherically
symmetric scalar field collapse \cite{Jalmuzna15}. The critical
solution in both 2+1-dimensional systems has a clear separation
between a contracting inner region where $M$ increases from $-1$ at
the regular center to a very small positive value at the boundary of
the shrinking central region, and an outer region, where $M$ remains
approximately constant and the matter is purely outgoing.

We have tentatively derived the type~II mass scaling law
$\delta=2\gamma$ from the observation that the mass in the atmosphere
of the quasistatic critical solution scales as $M(t)\sim -\Lambda
s(t)^2$, and the natural assumption that the apparent horizon mass is
given by this mass at the moment when the evolution leaves the
critical solution.

In summary, general relativity finds a way of making arbitrarily large
curvature and arbitrarily small black holes at the threshold of
collapse, even in 2+1 spacetime dimensions. It has to do
this in ingenious ways quite differently from 3+1 and higher
dimensions. Moreover, it does so very differently for the perfect fluid
with ultrarelativistic equation of state $P=\kappa\rho$ (for 
$\kappa \gtrsim 0.43$) and the massless scalar field \cite{Jalmuzna15}.

\begin{acknowledgements}
	The authors acknowledge the use of the IRIDIS 4 High Performance
	Computing Facility at the University of Southampton regarding the
	simulations that were performed as part of this work.
	
	Patrick Bourg was supported by an EPSRC Doctoral Training Grant to
	the University of Southampton.
\end{acknowledgements} 

\appendix
\section{No CSS solution for perfect fluid in $2+1$}
\label{appendix:appendixA}

We show here that in $2+1$ there are no nontrivial self-similar, spherically
symmetric perfect fluid solutions with a barotropic equation of state,
$P = \kappa \rho$, that is regular at the light cone.

We employ a similar notation as in Ref.~\cite{Hara97} namely, the independent
variables are defined by
\begin{equation}
x := -{r \over t}, \quad \tau := -\ln (-t)
\end{equation}
and
\begin{equation}
N := \frac{\alpha}{a x}, \, A := a^2, \, \omega := 4 \pi r^2 a^2 \rho.
\end{equation}
The equations of motion read
\begin{align}
{A' \over A} &= \frac{4 \omega (1+v^2 \kappa)}{x (1-v^2)},
\label{2p1:ApEqnReduced}\\
{\omega' \over \omega} &= \nonumber\\
& \hspace{-1cm} \frac{\left(1+v^2 \kappa\right) \left(\left(1-\kappa\right)
	\left(v^2 (2 \omega +1)+2 \omega \right)+2\right)}{x \left(1-v^2\right)
	\left(1-v^2 \kappa\right) }, \label{2p1:omegapEqnReduced}\\
{v' \over v} &= \frac{(1+v^2 \kappa) (1+2 \omega (1-\kappa))}{x (1-v^2 \kappa)}, \label{2p1:VpEqnReduced}\\
N &= -\frac{1+\kappa v^2}{v (1+\kappa)}. \label{2p1:NEqnReduced}
\end{align}
It is important to notice that the velocity is restricted to negative
values $-1 < v < 0$ and $A' \geq 0$, $\omega' \geq 0$, and $v' \leq 0$.

By definition, the light cone, $x=x_\text{lc}$, occurs at
$v^2(x_\text{lc}) = 1$.

First, we show that $x_\text{lc}$ is finite.
From Eq.~\eqref{2p1:VpEqnReduced}, we can find an upper bound for $|v|'$,
\begin{equation}
{|v|' \over |v|} \ge \frac{1+v^2 \kappa}{x (1-v^2 \kappa)}.
\end{equation}
The above inequality is separable and upon integration we find
\begin{equation}
x \leq \frac{C |v|}{1+\kappa v^2},
\end{equation}
where $C > 0$ is an integration constant. In particular,
we find an upper bound for $x_\text{lc}$,
\begin{equation}
x_\text{lc} \leq \frac{C}{1+\kappa} < \infty.
\end{equation}

Since $x_\text{lc}$ is finite, the ODE system needs to be regularized
at that point. Specifically, we must impose the numerators in
Eqs.~\eqref{2p1:ApEqnReduced} and \eqref{2p1:omegapEqnReduced} to vanish at
the light cone, which gives the constraint
\begin{equation}
\omega(x_{\text{lc}}) = 0.
\end{equation}
This constraint, together with the property that $\omega' \ge 0$, implies
\begin{equation}
\omega(x) = 0, \qquad x \in [0, x_\text{lc}],
\end{equation}
which proves our claim.

\section{Spherically symmetric static fluid}
\label{appendix:appendixB}

We review here the relevant properties of the static perfect fluid solutions
with $\Lambda \leq 0$. We refer the reader to Ref.~\cite{Carsten20} for a more
complete discussion.

For a given equation of state, these can
be parameterized by a length scale $s$, so that formally we can write
\begin{equation}
Z=\hat Z(R,s), \qquad Z:=\{R^2 \rho,a,\alpha,M\},
\end{equation}
where the hat denotes the static solution, when expressed in terms of
the radial coordinate $R$ and scale parameter $s$. (We come back to
the interpretation of $s$ below).

The functions $\hat Z(R,s)$ cannot be given in closed form, but the
quantities $Z$ and the area radius $R$ can be given explicitly in
terms of an auxiliary radial coordinate $y$ that is defined by
$\alpha=:y$.

We introduce the intermediate dimensionless quantities
\begin{equation}
\label{xdef}
x:={R\over s}
\end{equation}
and
\begin{equation}
\label{mudef}
\mu:=-\Lambda s^2.
\end{equation}
We can then write
\begin{equation}
\hat Z= \check Z(x,\mu), \qquad \check R = \sqrt{\mu} \ell x.
\end{equation}

We then have the following explicit expressions for the static solution,
but expressed in terms of the radial coordinate $y$ and parameter $\mu$: 
\begin{eqnarray}
\bar \alpha&=:&y, \\
\label{rhobar}
\bar\rho&=& {1-\mu \over 8 \pi \kappa \mu \ell^2}y^{-{1+\kappa\over\kappa}},\\
\bar a^{-1} &=& \mu y+(1-\mu)y^{-{1\over\kappa}}, \label{2p1:static:a}\\
\bar M&=&\bar x^2 \mu-\bar a^{-2}, \\
\bar x^2&=&\mu(y^2-1)+{2\kappa(1-\mu)\over1-\kappa}\left(1-y^{-{1-\kappa\over\kappa}}
\right). \nonumber \\
\label{2p1:static:x2} \\
\bar R &=& \sqrt{\mu} \ell \bar x.
\end{eqnarray}
The functions $\hat Z(R,s)$ are now given implicitly in terms of
$\bar R(y,\mu)$ and $\bar Z(y,\mu)$.

Note that only values $0\le\mu\le 1$
are physical, that $R=0$ is at $y=1$, and that (for $\mu>0$ only)
$R\to\infty$ as $y\to\infty$.

We also need to evaluate $\hat Z_{,R}$ and $\hat Z_{,s}$. For these, we
can derive the following expressions that are explicit in $y$ and $\mu$:
\begin{eqnarray}
\hat Z_{,s}&=&{2\over \ell\sqrt{\mu}}\left(\mu \bar Z_{,\mu}
-{\mu(\bar x^2)_{,\mu}+\bar x^2\over (\bar x^2)_{,y}}\bar Z_{,y}\right), \nonumber
\\ \\
\hat Z_{,R}&=&{2\bar x\over \ell\sqrt{\mu}(\bar x^2)_{,y}}\bar Z_{,y}.
\end{eqnarray}

For $\mu \ll 1$, we can distinguish a stellar interior and an
atmosphere, divided by a sharp turning point in $\hat \rho(R,s)$.
While the surface of the star is not defined
precisely in the presence of an atmosphere, we can take it to be at
\begin{equation}
\label{yc}
y = y_c := \bracket{1-\mu \over \kappa \mu}^{\kappa \over 1+\kappa},
\end{equation}
which marks both the maximum of $\bar a$ and the turning point of $\bar
x^2$, with respect to $y$.
Note that for $\mu \ll 1$, 
\begin{equation}
\bar x(y_c,\mu) =: \check x_c \simeq x_\star
\end{equation}

In the interior of the star we can neglect the first term in $\bar x^2$,
obtaining the approximate closed-form expression for $\check y$,
\begin{equation}
\check y\simeq \bracket{ 1- {1 \over 1-\mu}
	{x^2 \over x_\star^2}}^{-{\kappa \over 1-\kappa}},
\end{equation}
where we have defined
\begin{equation}
x_\star^2 := {2 \kappa \over 1-\kappa}. \label{2p1:static:xstar}
\end{equation}
Explicit approximate expressions for $\check Z$ then follow.

In the exterior of the star, we can approximate the second term in
$\bar x^2$ by its asymptotic value, obtaining
\begin{equation}
\label{yofxexterior}
\check y\simeq\mu^{-1}\left(-\check M_\infty +\mu x^2\right)^{1\over 2}
=\mu^{-1}\left(-\check M_\infty+{\check R^2 \over\ell^2}\right)^{1\over 2},
\end{equation}
where
\begin{equation}
\check M_\infty:={-(1+\kappa)\mu^2+2\kappa\mu\over 1-\kappa},
\label{2p1:static:Mtot}
\end{equation}
is the total mass of the system. This again gives explicit approximate
expressions for $\check Z$, in particular
\begin{equation}
\mu^2 \check \alpha^2\simeq \check a^{-2}\simeq -\check M_\infty+{\check R^2\over\ell^2}.
\end{equation}
We see that in the atmosphere, the metric is approximated by the BTZ
metric with fixed mass $\check M_\infty$ (and $t$ rescaled relative to the
convention for BTZ solutions), and so the fluid is approximated as a
test fluid, neglecting its self-gravity.

Substituting Eq.~\eqref{yofxexterior} into Eq.~\eqref{rhobar}, we find
\begin{equation}
\check \rho\simeq {1-\mu \over 8\pi\kappa \mu \ell^2}
\left(-{\check M_\infty\over \mu}+x^2\right)^{-{1+\kappa\over 2\kappa}}.
\end{equation}
The central density $\check \rho_0 := \check \rho(0,\mu)$ is related to
the total mass by
\begin{equation}
\label{2p1:Mtotrho0}
\lim\limits_{\mu \to 0} {\check M_\infty \check \rho_0 \over -\Lambda} = 
\frac{1}{4 \pi (1-\kappa)}.
\end{equation}

We now come back to the interpretation of $s$ as a length scale. For
$0<\mu\ll 1$, the surface $y=y_c$ is at $x\simeq x_\star$, and hence
at $R\simeq x_\star s$. In this sense, $x_\star s$ is the size of the
star. In the limiting case $\mu=0$, the star has a sharp surface at $x
= x_\star$, with $\check M_\infty=0$ in the vacuum exterior. The
exterior spatial geometry is then that of a cylinder of constant
radius. The limit $\mu \to 0$ is singular in the sense that
for vanishing $\mu$, the approximation $\check Z(x,\mu)\simeq \check
Z(x,0)$ is only valid for $x < x_\star$.

This means that in the limit where the size of the star is small
compared to the cosmological length scale $\ell$, $\mu=s^2/\ell^2\ll
1$, the family of static solutions becomes invariant under rescaling
$R$ and $\rho$ according to their dimensions, but only in the
interior of the star. In the atmosphere, $Z=\check Z(x,0)$ is not a
good approximation for small but finite $\mu$, and we need the full
form $Z=\check Z(x,\mu)$.

\section{The quasi-static solution}
\label{appendix:appendixC}

We model the critical solution as quasistatic, meaning that it
adiabatically goes through the sequence of static solutions, with $s$
now a function of $t$, and $|\dot s(t)|\ll 1$. We can then formally
expand the quantities $Z$ in even powers of $\dot s$, and $v$ in odd
powers.

In fact, from the exponential form of $s(t)$ [Eq.~\ref{st_ansatz}], it
follows that $\dot s=-\nu s/\ell=-\nu\sqrt{\mu}$, and hence $\dot
s^2=s\ddot s=\nu^2\mu$, and so the quasistatic approximation is
equivalent to the small-$\nu$ approximation. For now, however, we do
not assume the exponential form.

The leading and next order for $Z$ in the quasistatic ansatz are 
\begin{equation}
Z_\star(R,t)= Z_0(R,t)+\dot s^2(t)Z_2(R,t)+O(\dot s^4),
\end{equation}
where 
\begin{equation}
Z_0(R,t):= \hat Z(R,s(t)).
\end{equation}
As noted above, we do not have $\hat Z(R,s)$ in explicit form, only
$\bar Z(y,\mu)$. For the velocity, we make the ansatz
\begin{equation}
u_\star(R,t)=\dot s(t) u_1(R,t)+O(\dot s^3), 
\end{equation}
where we have defined
\begin{equation}
u:=\Gamma^2 v={v\over 1-v^2}.
\end{equation}
Clearly, for small $\dot s$ we have
\begin{equation}
v_\star(R,t)=\dot s(t) u_1+O(\dot s^3),
\end{equation}
but expanding $u$ rather than $v$ in a series in $\dot s$ makes sure that $|v|<1$.

To order $\dot s$, the Einstein equation (\ref{Mteqn}) becomes
\begin{equation} 
M_{{0},t} = \dot s \hat M_{,s} \simeq -16 \pi f_\Omega.
\end{equation}
This gives
\begin{equation}
\label{myu1}
\hat u_1:=-{\hat a \hat M_{,s} \over 16\pi (1+\kappa)R \hat \rho \hat \alpha}.
\end{equation}
As for $\hat Z$, we can compute $\bar u_1(y,\mu)$ explicitly, and as
expected, we find that $\check u_1(x,\mu)\simeq \check u_1(x,0)$ for $x
< x_\star$. However, $\check u_1(x,0)$ blows up at the surface
$x=x_\star$, while $\check u_1(x,\mu)$ is regular for all $x$.

As $R\to\infty$, $u_1$ goes to zero for $\kappa>1/2$,
approaches a constant value for $\kappa=1/2$, and diverges for
$\kappa< 1/2$. Hence, the expansion in $\dot s$ breaks down in the
atmosphere for $\kappa<1/2$ at sufficiently large radius $R$
and contraction speed $|\dot s|$. 
However, for the values of $\dot s$ for which we show plots, this is
not a problem for values of $R$ on our numerical grid.

The momentum balance law is obeyed to leading order in $\dot
s^2$ by construction. Going to the next order, we see that the
$Y_{,t}$ term, with $Y\propto u_1\dot s$, produces a term proportional
to $\ddot s$.We consider $s\ddot s$ as the same order as $\dot s^2$,
which is true when $s(t)$ is either exponential or a power.

The ansatz
\begin{equation}
s\ddot s=G(\mu)\dot s^2
\end{equation}
results in an inhomogeneous first-order ODE in $y$ for $\rho_2$,
with $\mu$ merely a parameter, and no explicit appearance of $s(t)$ or
$t$ derivatives, and so can be thought of as a separation of variables ansatz.

This ODE in $y$ for $\rho_2$ contains $a_2$, although $\alpha_2$
drops out when we use the background momentum balance law. To close the
system, we must perturb therefore only the Einstein equation
(\ref{dMdr}) to $\mathcal{O}(\dot s^2)$ to get an ODE for $a_2$.

The resulting system of two inhomogeneous first-order ODEs in $y$ for
$\rho_2$ and $a_2$ can also be rewritten as a single second-order
inhomogeneous ODE for $M_2$. Obviously, the solutions of the corresponding
homogeneous ODE, obtained by setting $u_1=0$, are simply the static perturbations
of the regular static solution. One of these is singular at the origin
and so is ruled out by regularity. The other, with $M_2=0$ at the
origin and finite at infinity, is the infinitesimal change
$\hat M_{,s}$ from one regular static solution to a neighboring one.

We had hoped to find a natural boundary condition for $M_2$ at
infinity that would select the value $G(\mu)=1$ of the separation
constant, in order to predict the observed exponential form of
$s(t)$. This would have been similar in spirit to the approach of
Ref.~\cite{Bizon04} for a quasistatic critical solution (in Yang-Mills on
flat spacetime in 4+1 dimensions).

However, we have not found such a boundary condition. In particular,
$M_2$ is finite at $R=\infty$ (and can then be set to zero there by
adding a multiple of $\hat M_{,s}$) for all $\kappa\ge 1/2$, but for
all $\kappa<1/3$, it blows up at $R=\infty$, with $G(\mu)$ only
affecting lower-order terms. For $1/3 < \kappa < 1/2$, $M_2$ is finite,
but since $u_1$ is not, the quasistatic ansatz is also not valid in
this case, as $\dot s u_1$ can then not be considered as small. 

\section{Stationary test fluid solutions}
\label{appendix:appendixD}

To understand better what happens for $\kappa \leq 1/2$, where $u_1$
diverges at infinity, we note that the fluid in the atmosphere can be
approximated as a test fluid on a fixed BTZ spacetime with mass
$\check M_\infty$.

The solutions describing a stationary test fluid with constant mass
flux on a BTZ spacetime with mass $M$ can be given in implicit form as
\begin{eqnarray}
\label{constantflux_veqn}
v(1-v^2)^{1-\kappa\over 2\kappa}&=&{f_\Omega \over C \rho_0 (1+\kappa)}
{(-M-\Lambda R^2)^{1-\kappa\over 2\kappa}\over R}\nonumber\\ \\
\rho&=&\rho_0(-M-\Lambda R^2)^{-{1+\kappa\over 2\kappa}}
(1-v^2)^{1+\kappa\over 2\kappa}.\nonumber\\
\end{eqnarray}
Here, the free parameter $f_\Omega$ is the constant mass flux, $C := a
\alpha$ is a constant in vacuum that depends on the normalization of
the time coordinate $t$, and the free parameter $\rho_0$ is an overall
factor in $\rho$ chosen so that it is the density at the center in the
static solution on AdS spacetime.

We note that generically, Eq.~\eqref{constantflux_veqn} has either two
solutions $v$ or none. From
\begin{equation}
\label{const_flux_v_asymptotic}
v(1-v^2)^{1-\kappa\over 2\kappa} \sim R^{1-2\kappa\over\kappa},
\end{equation}
we see that for $\kappa > 1/2$, either $v\to 0$ or $v \to \pm 1$ as
$R\to\infty$. The $v\to 1$ solution is the one relevant for our
critical solution. We then have
\begin{equation}
\rho \sim R^{-{1+\kappa\over 1-\kappa}}.
\end{equation}
However, for $\kappa < 1/2$, the right-hand side of
Eq.~\eqref{constantflux_veqn} increases with $R$, and so $v$ is defined
only up to some maximum value of $R$, beyond which the constant flux
solution does not exist (for given $f_\Omega$ and $\rho_0$).

In the top two plots of
Fig.~\ref{fig:vs2rho_constant_flux_kappa047_and_054}, we show that
this ansatz for $\kappa=0.47$ is in good agreement with our numerical
solution up to the numerical outer boundary. The radius after which
$v$ is not defined as explained above lies outside our numerical grid.


\end{document}